\documentclass{aa}  

\usepackage{graphicx}
\usepackage{epstopdf}
\usepackage{txfonts}
\usepackage{color}
\usepackage{booktabs}
\usepackage{multirow}
\usepackage{tabularx}
\definecolor{linka}{rgb}{0.0,0.0,0.7}
\definecolor{linkb}{rgb}{0.0,0.0,0.5}
\definecolor{linkc}{rgb}{0.0,0.0,0.3}
\definecolor{cmt}{rgb}{0.5,0.0,0.0}
\definecolor{al}{rgb}{0.6,0.2,0.0}
\definecolor{mvn}{rgb}{0.4,0.4,0.0}
\definecolor{corr}{rgb}{0,0.0,0.4}
\definecolor{ref}{rgb}{0.4,0.0,0.0}
\definecolor{idea}{rgb}{0.4,0.0,0.0}
\usepackage[colorlinks=true,linkcolor=linkb,citecolor=linkc,filecolor=linka,urlcolor=linka,bookmarks=true,bookmarksopen=true,bookmarksopenlevel=3,plainpages=false,pdfpagelabels=true]{hyperref}
\usepackage{natbib}
\graphicspath{{./}{../figures/}{./figures/}}

\newcommand{\hminus}{H$^-$}

\newcommand{\fei}{Fe\,\textsc{i}}
\newcommand{\tii}{Ti\,\textsc{i}}

\newcommand{\imax}{{IMaX}}

\newcommand{\sunrise}{\textsc{Sunrise}}

\newcommand{\mgeff}{g_{\mbox{\tiny{eff}}}}
\newcommand{\tgeff}{$\mgeff$}
\newcommand{\mvmax}{V_{\mbox{\tiny{max}}}}
\newcommand{\tvmax}{$\mvmax$}

\newcommand{\fig}[1]{Fig.~\ref{#1}} 
\newcommand{\Fig}[1]{Figure~\ref{#1}} 
\newcommand{\eq}[1]{Eq.~\ref{#1}} 
\newcommand{\figs}[1]{Figs.~\ref{#1}} 
\newcommand{\tab}[1]{Tab.~\ref{#1}} 
\newcommand{\Tab}[1]{Table~\ref{#1}} 
\newcommand{\sect}[1]{Sec.~\ref{#1}} 
\newcommand{\Sect}[1]{Section~\ref{#1}}

\newcommand{\xy}[2]{$x$=#1\arcsec{}, $y$=#2\arcsec{}}

\newcommand{\fea}{Fe\,\textsc{i}\,15648.5\,\AA{}}
\newcommand{\feb}{Fe\,\textsc{i}\,15652.9\,\AA{}}
\newcommand{\feahin}{Fe\,\textsc{i}\,6302.5\,\AA{}}
\newcommand{\febhin}{Fe\,\textsc{i}\,6301.5\,\AA{}}

\newcommand{\colfig}[3][1.]{\begin{figure}\centering
    \includegraphics[width=#1\linewidth,clip=TRUE]{#2}
    \caption{#3}
    \label{#2}
\end{figure}}

\begin{document}

\title{Probing deep photospheric layers of the quiet Sun with high magnetic sensitivity}
\titlerunning{High magnetic sensitivity probing of the quiet Sun}
\authorrunning{A. Lagg et al.}

\author{A. Lagg
  \inst{1}
  \and
  S.~K. Solanki\inst{1,5} 
  \and
  H.-P. Doerr\inst{1} 
  \and
  M.~J.Mart\'inez Gonz\'alez\inst{2,9} 
  \and
  T. Riethm\"uller\inst{1} 
  \and
  M. Collados Vera\inst{2,9}
  \and
  R. Schlichenmaier\inst{3} 
  \and
  D. Orozco Su\'arez\inst{10} 
  \and
  M. Franz\inst{3} 
  \and
  A. Feller\inst{1} 
  \and
  C. Kuckein\inst{4} 
  \and
  W. Schmidt\inst{3} 
  \and
  A. Asensio Ramos\inst{2,9} 
  \and
  A. Pastor Yabar\inst{2,9} 
  \and
  O. von der L\"uhe\inst{3}
  \and
  C. Denker\inst{4} 
  \and
  H. Balthasar\inst{4} 
  \and
  R. Volkmer\inst{3} 
  \and
  J. Staude\inst{4} 
  \and
  A. Hofmann\inst{4} 
  \and
  K. Strassmeier\inst{4} 
  \and
  F. Kneer\inst{6} 
  \and
  T. Waldmann\inst{3} 
  \and
  J.~M. Borrero\inst{3} 
  \and
  M. Sobotka\inst{7} 
  \and
  M. Verma\inst{4} 
  \and
  R.~E. Louis\inst{4}
  \and
  R. Rezaei\inst{2} 
  \and
  D. Soltau\inst{3}
  \and
  T. Berkefeld\inst{3}
  \and
  M. Sigwarth\inst{3}
  \and
  D. Schmidt\inst{8}
  \and
  C. Kiess\inst{3}
  \and
  H. Nicklas\inst{6}
}

\institute{Max-Planck-Institut f\"ur Sonnensystemforschung, Justus-von-Liebig-Weg 3, 37077 G\"ottingen, Germany
\and
Instituto de Astrof\'isica de Canarias, C/ V\'ia L\'actea, La Laguna, Spain
\and
Kiepenheuer-Institut f\"ur Sonnenphysik, Sch\"oneckstr.~6, 79104 Freiburg, Germany
\and
Leibniz-Institut f\"ur Astrophysik Potsdam, An der Sternwarte 16, 14482 Potsdam, Germany
\and
School of Space Research,
Kyung Hee University, Yongin, Gyeonggi 446-701, Republic of Korea
\and
Institut f\"ur Astrophysik, Friedrich Hund Platz 1, 37077 G\"ottingen, Germany
\and
Astronomical Institute, Academy of Sciences of the Czech Republic, Fri\v{c}ova 298, 25165 Ond\v{r}ejov, Czech Republic
\and
National Solar Observatory, 3010 Coronal Loop, Sunspot, NM 88349, USA
\and
Dept. Astrof\'isica, Universidad de La Laguna, E-38205, La Laguna, Tenerife, Spain
\and
Instituto de Astrof\'isica de Andaluc\'ia (CSIC), Glorieta de la Astronom\'ia, 18008 Granada, Spain
\\ \email{lagg@mps.mpg.de}}
\offprints{A. Lagg, \email{lagg@mps.mpg.de}}
\date{accepted: April 29, 2016}

\abstract
{Investigations of the magnetism of the quiet Sun are hindered by extremely weak polarization signals in Fraunhofer spectral lines. Photon noise, straylight, and the systematically different sensitivity of the Zeeman effect to longitudinal and transversal magnetic fields result in controversial results in terms of the strength and angular distribution of the magnetic field vector.}
{The information content of Stokes measurements close to the diffraction limit of the $1.5$\,m GREGOR telescope is analyzed. We took the effects of spatial straylight and photon noise into account.}
{Highly sensitive full Stokes measurements of a quiet-Sun region at disk center in the deep photospheric \fei{} lines in the 1.56\,$\mu$m region were obtained with the infrared spectropolarimeter GRIS at the GREGOR telescope. Noise statistics and Stokes $V$ asymmetries were analyzed and compared to a similar data set of the Hinode spectropolarimeter (SOT/SP). Simple diagnostics based directly on the shape and strength of the profiles were applied to the GRIS data. We made use of the magnetic line ratio technique, which was tested against realistic magneto-hydrodynamic simulations (MURaM).}
{About 80\% of the GRIS spectra of a very quiet solar region show polarimetric signals above a 3$\sigma$ level. Area and amplitude asymmetries agree well with small-scale surface dynamo magnetohydrodynamic simulations. The magnetic line ratio analysis reveals ubiquitous magnetic regions in the ten to hundred Gauss range with some concentrations of kilo-Gauss fields.}
{The GRIS spectropolarimetric data at a spatial resolution of $\approx$0\farcs4 are so far unique in the combination of high spatial resolution scans and high magnetic field sensitivity. Nevertheless, the unavoidable effect of spatial straylight and the resulting dilution of the weak Stokes profiles means that inversion techniques still bear a high risk of misinterpretating the data.}

\maketitle

\section{Introduction}

Even during the maximum of solar activity, 90\% of the solar photosphere is covered by the so-called quiet Sun \citep{sanchezalmeida:04a}. This term refers to regions of undisturbed granular convection patterns with no significant polarization signals in traditional synoptic magnetograms. Describing the magnetism of these regions is therefore crucial to understand not only how the Sun produces these small-scale magnetic fields, but also how the magnetic energy is transported to higher atmospheric layers. Unfortunately, characterizing the quiet-Sun magnetism from measurements is extremely difficult: Small-scale magnetic fields produce only weak signals that vary on granular timescales of minutes, which presents a challenge even for the most advanced spectropolarimeters.

The characterization of the quiet-Sun magnetic fields relies on the interpretation of the polarization signals in Fraunhofer lines, either altered by the Hanle effect \citep[for a review see, e.g.,][]{stenflo:11} or produced by the Zeeman effect \citep[e.g.,][]{sanchezalmeida:11}. The first has the advantage of being unaffected by signal cancellation in turbulent regimes, the latter is easier to interpret. In this work we treat the polarization signal in highly magnetically sensitive photospheric lines that is induced by the Zeeman effect. A pioneering work by  \citet{stenflo:73} revealed strong kilo-Gauss magnetic field concentrations on subarcsecond scales as a signature of convective field intensifications \citep{grossmanndoerth:98a,danilovic:10a,lagg:10b,requerey:14}. Using the magnetic line ratio (MLR) technique, \citeauthor{stenflo:73} found these fields to be surrounded by weak fields in the range of a few Gauss. At least since then, a controversial debate about the distribution of field strengths and inclination in quiet-Sun areas has begun, with contradicting results that are sometimes even based on data sets from the same instruments \citep[e.g., from Hinode SOT/SP:][]{orozco:07b,lites:08a,jin:09,asensioramos:09,martinezgonzalez:10a,stenflo:10,borrero:11,ishikawa:11a,orozco:12a,bellotrubio:12a,stenflo:13b,asensioramos:14a}, or when using different spectral lines for the analysis \citep[e.g., \fei{} lines at 1.56\,$\mu$m: ][]{stenflo:87a,lin:95,solanki:96b,meunier:98,khomenko:03,khomenko:05a,martinezgonzalez:08b,beck:09}. Several reviews on this topic are available \citep[e.g.,][]{dewijn:09,steiner:12,borrero:15a}.

This controversy arises because the Stokes signals are difficult to interpret. The weak signals are often barely above the noise level, requiring the setting of thresholds above which the signal is believed to be trustworthy. Unfortunately, the different sensitivity of the Zeeman effect to transverse and longitudinal magnetic fields makes a bias-free definition of such thresholds almost impossible. The main tool for retrieving atmospheric parameters is by inverting the radiative transfer equation, and this therefore may deliver biased results for such low-signal profiles. Additional complexity stems from the small-scale nature of these internetwork fields. Atmospheric seeing, straylight, and the point spread function (PSF) of the telescope dilute the weak signals even further, leading to ambiguous interpretations of magnetic field strength, inclination, and fill fraction.


Progress in the field of quiet-Sun magnetism therefore requires both highest magnetic sensitivity and highest spatial resolution. The GREGOR Infrared Spectrograph \citep[GRIS,][]{collados:07,collados:12}, in scientific operation since 2014, opens a new domain by combining these two attributes. The \fea{} line with a Land\'e factor of $g=3$ is, mainly because of the wavelength dependence of the Zeeman effect, approximately three times more sensitive to magnetic fields than the widely used \fei{} lines at 6302.5\,\AA{} ($g=2.5$), 6173.3\,\AA{} ($g=2.5$), and 5250.2\,\AA{} ($g=3$). The noise level of GRIS can be as low as $\approx2\times10^{-4}$ of the continuum intensity. And finally, the large aperture of the GREGOR telescope of 1.5\,m brings the spatial resolution in the infrared to values previously reserved for observations in the visible.

In this paper we present a GRIS scan of a highly undisturbed quiet-Sun area (\sect{data}). With techniques based on analyzing the shape and strength of Stokes profiles, without involving sophisticated modeling, we try to circumvent the above-mentioned caveats in the interpretation of Stokes profiles. We compare the GRIS data with a deep-magnetogram mode scan obtained with Hinode SOT/SP and with magneto-hydrodynamic simulations. Noise statistics are presented in \sect{noise} and are followed by the analysis of the complexity of the Stokes profiles using area and amplitude asymmetries (\sect{complexity}). In \sect{mlr} we compute the MLR and the ratio between linear and circular polarization (LP$/$CP) from the Stokes profiles to infer the magnetic field strengths and obtain some information about the magnetic field inclination in the GRIS data in a nearly model-independent manner and compare it to radiation magneto-hydrodynamic (MHD) simulations. \Sect{conclusion} summarizes our findings.

\section{Data sets\label{data}}

\subsection{GRIS observations\label{grisobs}}

On 17 September 2015 we observed a quiet-Sun region very close to disk center (solar coordinates $x=10$\arcsec{}, $y=-3$\arcsec{}, $\mu=\cos\Theta=1.00$) with GRIS mounted at the 1.5\,m GREGOR telescope \citep{schmidt:12,soltau:12}. The region was selected using SDO/HMI magnetograms \citep{schou:12} with the goal of avoiding magnetic network flux concentrations as much as possible. The quiet Sun was scanned by moving the 0\farcs135 wide slit with a step size of 0\farcs135 over a 13\farcs5 wide region. The pixel size along the slit is also 0\farcs135, corresponding to half the diffraction-limited resolution of GREGOR at the observed wavelength. The total exposure time per slit position was 4.8\,s, accumulated from 20 camera readouts with 60\,ms exposure time per polarimetric modulation state. The total duration of the scan was from 08:26~UT until 08:40~UT.

The standard data reduction software was applied to the GRIS data for dark current removal, flat-fielding, polarimetric calibration, and intensity to Stokes $Q$, $U$, and $V$ cross-talk removal. The polarimetric calibration was performed using the built-in calibration unit in F2 \citep{hofmann:12} in combination with a model for the telescope polarization. The modulation efficiencies for $Q$, $U$, and $V$ were $\approx$0.54, which is close to the theoretical limit of $1/\sqrt{\smash[b]{3}}$.

The seeing conditions during the scan were very good, allowing us to produce images close to the diffraction limit with the blue-imaging-channel of the GREGOR Fabry-P\'erot Interferometer \citep[GFPI,][]{puschmann:12a} operated in the blue continuum. However, the residual motion of the image over the time span needed to record one slit position reduced the spatial resolution of the continuum image assembled from the individual GRIS scans to 0\farcs40 (i.e., almost at the angular resolution defined by the Rayleigh criterion at 1.565\,$\mu$m of 0\farcs26). This is close to the resolution achievable with the Hinode spectropolarimeter (0\farcs32) and superior to previous studies in these spectral lines  \citep[e.g., 1\arcsec{} in][]{lin:99,khomenko:03,martinezgonzalez:08b}. The root mean square contrast of the continuum image is 2.3\%.

The observed spectral region covers a 40.5\,\AA{} wide window around 1.56\,$\mu$m, sampled at 40.1\,m\AA{}$/$pixel. In this wavelength region, the \hminus{} opacity has a minimum that allows an unobstructed view to deep photospheric layers \citep[see][]{borrero:16a}. Of the several available \fei{} lines in this region, we selected the well-known \fei{} line pair at 15648.5\,\AA{} and 15652.9\,\AA{}. With a  Land\'e factor of $g=3,$ the first line shows the highest Zeeman sensitivity of all unblended spectral lines in the visible and the near-infrared range, matched only by the $g=2.5$ \tii{} line at 2.231\,$\mu$m \citep{ruedi:98a}. The second line (effective Land\'e factor $\mgeff=1.53$) is formed under very similar atmospheric conditions, making this line pair well suited (although not completely ideal; see \sect{mlr}) for analysis methods such as the MLR technique we present below. The spectral resolution of the GRIS dataset was determined by comparison with the FTS spectral atlas of \citet{livingston:91} to be $\lambda/\Delta\lambda \approx 110\,000$, with an unpolarized spectral straylight contribution of 12\%.


A low noise level is of crucial importance for a proper analysis of the magnetic signatures in quiet-Sun areas. With the above-mentioned values for exposure time, spectral, and spatial sampling, we achieved a noise level of $4\times10^{-4}$ in the continuum wavelength points in Stokes $Q$, $U$, and $V$ for every pixel in the map. We applied the following two techniques to further decrease this noise level: (1) Spatial binning: The spatial resolution was limited by the seeing conditions to 0\farcs40. A spatial sampling of 0\farcs20 is therefore sufficient to preserve the entire information contained in the observations. A rebinning of the maps from originally 0\farcs135$/$pixel to 0\farcs20$/$pixel, based on fast Fourier transformations, increased the number of photons per pixel by a factor of $(0.2/0.135)^2$ and decreased the noise level to $\approx$3$\times10^{-4}$. (2) Spectral binning: A sampling of 80\,m\AA{}$/$pixel is sufficient for the spectral resolution of the GRIS scans. This was achieved by binning together two pixels in the spectral direction, which resulted in a further reduction of the noise level to $\sigma\approx2.2\times10^{-4}$ of the continuum intensity. The reduction of the noise level by this spatial and spectral binning is slightly lower than predicted from the plain photon statistics, indicating that systematic effects (e.g., detector readout noise, spatial or spectral fringing, seeing-induced cross-talk) start to play a role \citep[see also][]{franz:16a}.

\colfig{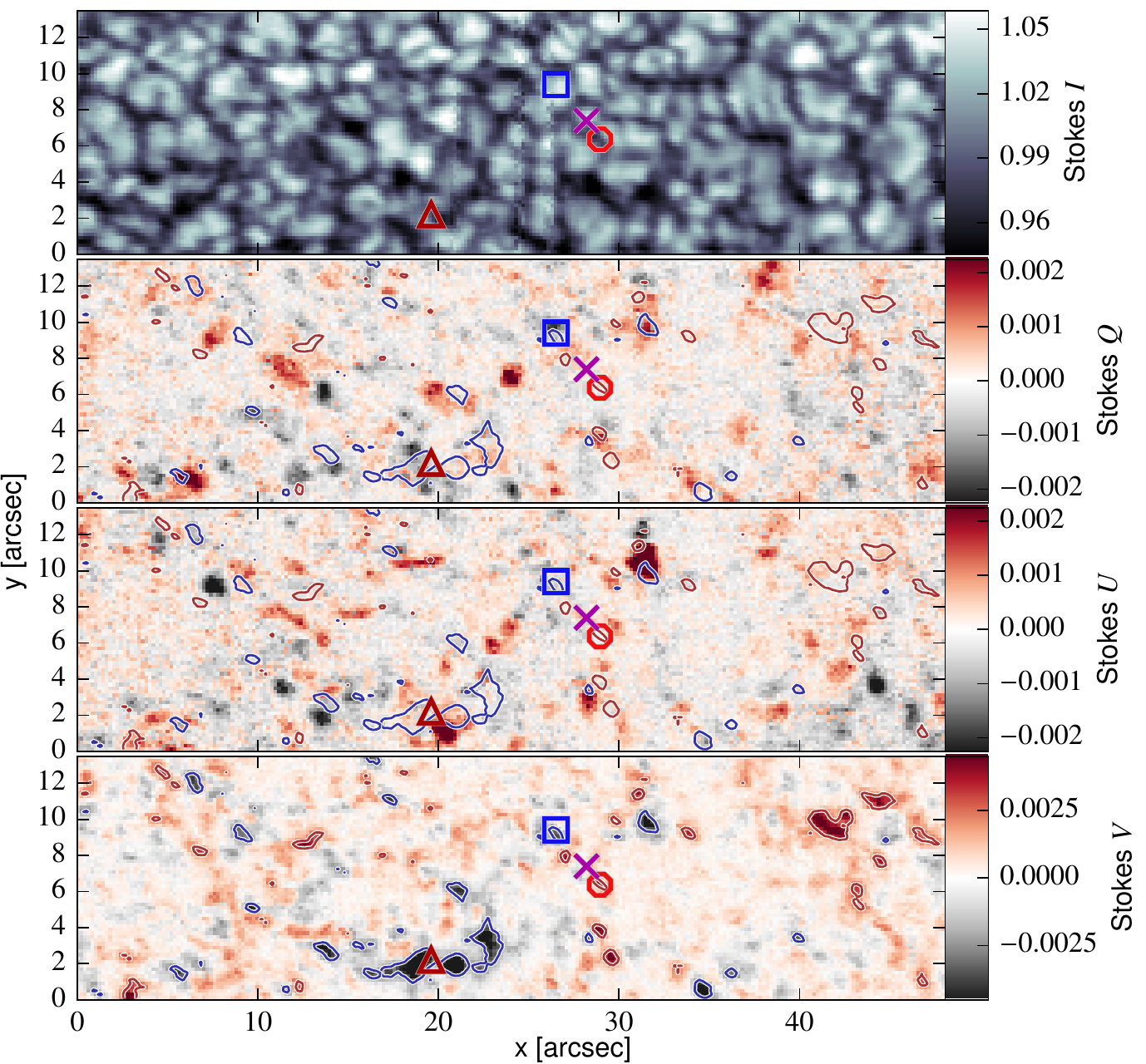}{Stokes maps assembled from the GRIS scan of the quiet-Sun region at the disk center. The top panel shows the continuum intensity at 1.56\,$\mu$m, the $Q$, $U$, and $V$ maps are averages over a 0.75\,\AA{} wide spectral window using the nominal wavelength of the \fea{} line as the central position for $Q$ and $U$, and as the lower limit for $V$. The abscissa $x$ is the slit direction, the ordinate $y$ is the scan direction. The blue and red contours mark the Stokes $V$ levels with $|V|=0.002$. The symbols mark the location of sample profiles, shown later in \figs{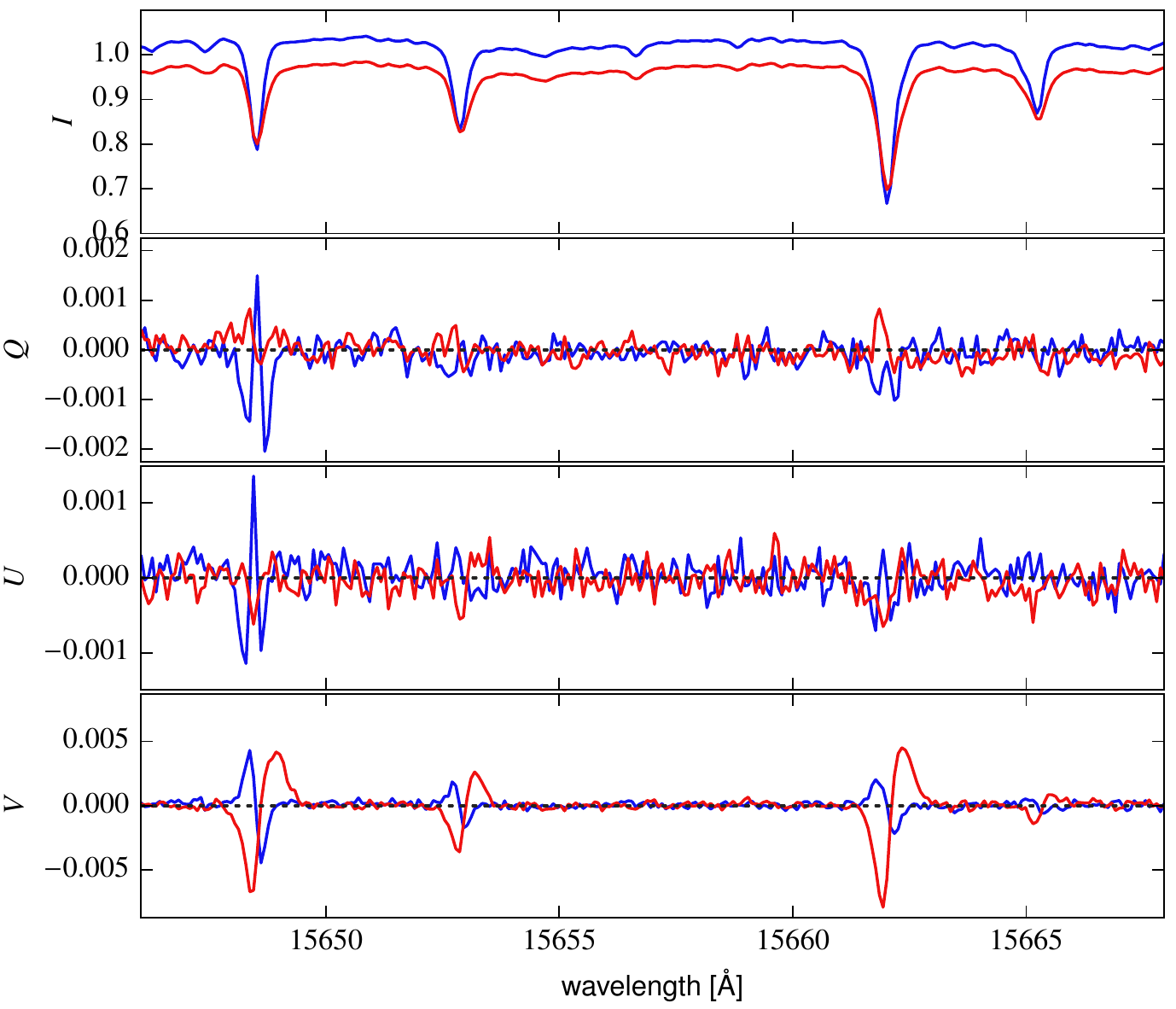}, \ref{prof_gris_mhd_kG.pdf}, and \ref{prof_gris_mhd_hG.pdf}. The contours outline a Stokes $V$ value of $\pm0.0025$.}

The quiet-Sun scan obtained with GRIS after this processing is presented in \fig{Stokes-gris.pdf}. Despite the long integration time per slit position and the long wavelength of the observation in the infrared, the intensity map (top panel) shows remarkable details in the granulation pattern.  The Stokes $Q,U,V$ parameters (lower three panels) show signals well above the noise level nearly everywhere in the observed region. The contours of the Stokes $V$ signal in the linear polarization $Q$ and $U$ maps show that the correlation between the linear polarization (LP) and the circular polarization (CP) is very low. This demonstrates that the patches containing horizontal magnetic fields do not coincide with the vertical flux tubes. Some of the strong CP features are connected through LP patches, indicative of small-scale magnetic loops (see, e.g., \xy{32}{10\,--\,12} in \fig{Stokes-gris.pdf}). A more detailed discussion about these connections can be found in \citet{martinezgonzalez:16a}.

A slight increase in the noise level of $\approx$15\% is shown in the left- and rightmost part of the Stokes $Q$ and $U$ maps. This is the result of not compensating for the image rotation during the observation \citep{volkmer:12}. In combination with the temporal modulation, this rotation introduces an increasing amount of cross-talk at a level of $2\times10^{-5}$ with increasing distance from the center of the rotation, which is the central position of the slit. An image derotator, available at GREGOR in the 2016 observing season, will avoid this problem in the future. The vertical stripes in the continuum intensity map around $x\approx30$\arcsec{} are a result of dust grains on the spectrograph slit. The apparent distortions parallel to the slit direction (i.e., the $x$-axis) are unavoidable effects of the seeing because the quality of the image correction by the GREGOR adaptive optics system \citep[GAOS,][]{berkefeld:12} decreases with distance from the lock point, which was centered on the spectrograph slit.

\colfig{grisprofGR-IG.pdf}{GRIS Stokes profile in a granule (blue, indicated with the blue square in \fig{Stokes-gris.pdf}) and intergranular lane (red, indicated with the red circle in \fig{Stokes-gris.pdf}).}

The importance of combining the low polarimetric noise level and highly magnetically sensitive spectral lines to correctly interpret the magnetic properties is illustrated by individual Stokes profiles. \Fig{grisprofGR-IG.pdf} shows a part of the spectral region for a profile in an intergranular lane (red) and in the center of a granule (blue). The positions of these profiles are indicated in \fig{Stokes-gris.pdf} with the red circle and the blue square. Both profiles show Stokes $V$ signals well above the noise level of 2$\times10^{-4}$ in Stokes $V$, indicating the presence of a significant line-of-sight magnetic field component\footnote{Since the observed region is located at disk center, the line-of-sight direction is identical with the surface normal.}. The strength of these signals is sufficient to be detectable for standard spectropolarimeters, which is not the case for the Stokes $Q$ and $U$ signals: here only the \fea{} line shows a clear response to the horizontal component of the magnetic field, obviously present in the center of the granule. Without the low noise level and the high magnetic sensitivity, it is impossible to determine the correct strength and orientation of the magnetic field vector for such a low-flux feature.

\subsection{Hinode SOT/SP observations}

The high-quality data from the spectropolarimeter onboard the Hinode spacecraft \citep[Hinode SOT/SP,][]{kosugi:07a,tsuneta:08a,suematsu:08a,ichimoto:08a} represented a major step forward in understanding quiet-Sun magnetism at the time of the launch of Hinode. SOT/SP scans serve as an observational benchmark for investigations of quiet-Sun magnetism. To evaluate the quality of the GRIS observations, we compared data from the two instruments. Ideally, the comparison should be done by scanning the same region on the Sun simultaneously with both instruments. Unfortunately, the long-exposure SOT/SP scans with high magnetic sensitivity were only available in the early phase of the Hinode mission, allowing only for a statistical comparison between Hinode and GRIS results.

\colfig{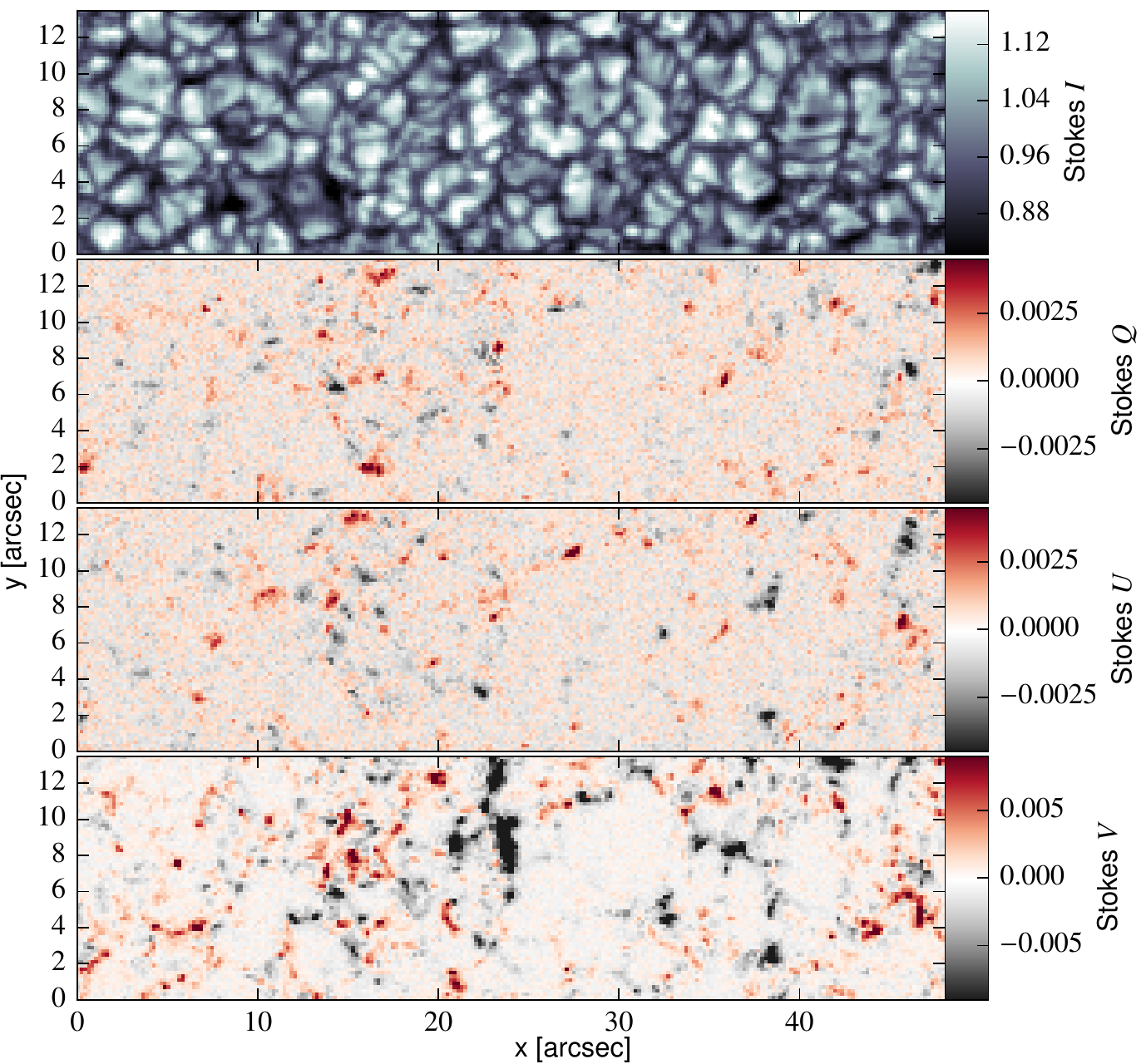}{Same as \fig{Stokes-gris.pdf}, but for a different quiet-Sun area, scanned with Hinode SOT/SP. Here the reference wavelength for the $Q$, $U$, and $V$ maps is the nominal position of the \fei{} 6302.5\,\AA{} line. The data set was clipped to the same size as the GRIS scan.}

For this comparison we selected a Hinode SOT/SP scan with the lowest possible noise level. On 23 September 2007, 08:16\,--\,09:25~UT, Hinode observed a quiet region in the so-called  deep magnetogram mode, which means by integrating over many rotations of the modulator. The data were reduced using the standard SOT/SP data reduction software \citep{lites:13a,lites:13b} based on \textsl{Solar Software} \citep{freeland:98}. The total exposure time of 12.8\,s per slit position of this 50\arcsec{} wide scan at disk center resulted in a noise level of $\sigma\approx6.9\times10^{-4}$ of the continuum level for the unbinned Stokes $Q,U,V$ profiles. The modulation efficiencies for Stokes $Q$, $U$, and $V$ are approximately 0.51 \citep{tsuneta:08a}. The SOT/SP data were FFT-resampled to a pixel size of 0\farcs20 to enable a 1:1 comparison with the GRIS data. This resampling reduced the noise level to 5.9$\times10^{-4}$. Since the SOT/SP data are critically sampled in the spectral direction, no binning in wavelength was applied. The resulting Stokes maps are displayed in \fig{Stokes-sotsp128.pdf}.


The Stokes signals in the SOT/SP maps show on average higher amplitudes than in the GRIS maps (note the different scaling for the two figures). At the same time, the magnetic features in the SOT/SP maps exhibit a smaller spatial extension. Both effects are most likely a result of the seeing-free conditions in space, which prevents the dilution of the Stokes signals by the broader wings of the spatial PSF of the GREGOR telescope.

\subsection{MURaM simulations\label{mhddata}}

For comparison with the GRIS scan we make use of two snapshots produced with the MURaM code \citep{voegler:05a}. The first snapshot is a non-gray version of the run O16bM described in \citet{rempel:14a} with a horizontal and vertical resolution of 16\,km. It is a small-scale dynamo run (hereafter referred to as MHD/SSD) with an open bottom boundary, allowing for the upflow of the horizontal magnetic field that emulates the presence of a deep, magnetized convection zone. The same snapshot was analyzed recently in \citet{danilovic:16a}, who compared various observables between the snapshot and Hinode SOT/SP observations. The cube contains almost exclusively weak fields in the range from below 10 to a few hundred Gauss at optical depth unity, with tiny kilo-Gauss field concentrations in a few coalescent intergranular lanes.

The second MHD snapshot was also calculated with the non-gray version of the MURaM code. The simulation box is 32.6$\times$32.6\,Mm$^2$ in its horizontal dimensions and has a depth of 6.1\,Mm. The cell size of the simulation is 40\,km in the two horizontal directions and 16\,km in the vertical direction. Stokes profiles measured with the \imax{} instrument flown on the \sunrise{} balloon-borne observatory \citep{solanki:10a,martinezpillet:11,barthol:11}  were used to determine the initial conditions of the atmospheric stratification in the cube \citep{riethmueller:16a}. The simulation was run for two hours of solar time to reach a statistically relaxed state. The boundary conditions were periodic in the horizontal directions and closed at the top boundary of the box. A free in- and outflow of plasma was allowed at the bottom boundary under the constraint of total mass conservation. The $\tau=1$ surface for the continuum at 500\,nm was on average reached about 700\,km below the upper boundary. This cube contains small sunspots, pores, and plage with magnetic field strengths up to 3\,kG at optical depth unity and is devoid of completely quiet solar regions, just like AR~11768 that was observed by \imax{} on 12 June 2013 at 23:40~UT. We refer to this MHD run as MHD/IMaX.

For the analysis in this paper we used the forward module of the SPINOR code \citep{solanki:87a,frutiger:00a,frutiger:thesis} to compute spectra in several \fei{} lines in the 1.56\,$\mu$m region, including the \fea{} and \feb{} lines. The data were spatially degraded by applying a PSF corresponding to the theoretical GREGOR PSF (calculated from aperture, central obscuration, and spider), which was additionally broadened by a Gaussian with a full-width at half-maximum (FWHM) of 0\farcs25 to match the spatial resolution of the GRIS scan of 0\farcs40. A Lorentzian was added with a width of 0\farcs75 and an amplitude of 0.05 to mimic the spatial straylight of GREGOR. With this PSF the root mean square contrast of the continuum intensity was reduced from 9\% in the original MHD cube to match the observed contrast of 2.3\% in the GRIS data. In addition, the histogram of the continuum intensity between the GRIS scan and the MHD data after this degradation agreed well. The data from the MHD cube were rebinned to match the pixel size of the GRIS observations (0\farcs20). It should be noted that after this spatial degradation, approximately 80\% of the photons originating from the 1:1 mapped solar area of a single pixel end up in the surrounding pixels of the detector. A spectral degradation with a Gaussian with 150\,m\AA{} FWHM and an added unpolarized spectral straylight component of 12\%, matching the values determined from the GRIS scan, completed the degradation.





\section{Noise level\label{noise}}

The GRIS scan of the quiet-Sun region offers an unprecedented combination of spatial resolution and polarimetric accuracy. Together with the high Zeeman-sensitivity of the \fea{} line, the detection and characterization of the weak signals from small-scale magnetic fields are pushed toward a new limit. This can be demonstrated by comparing the noise statistics of the GRIS data to the SOT/SP deep-magnetogram scan.

\colfig{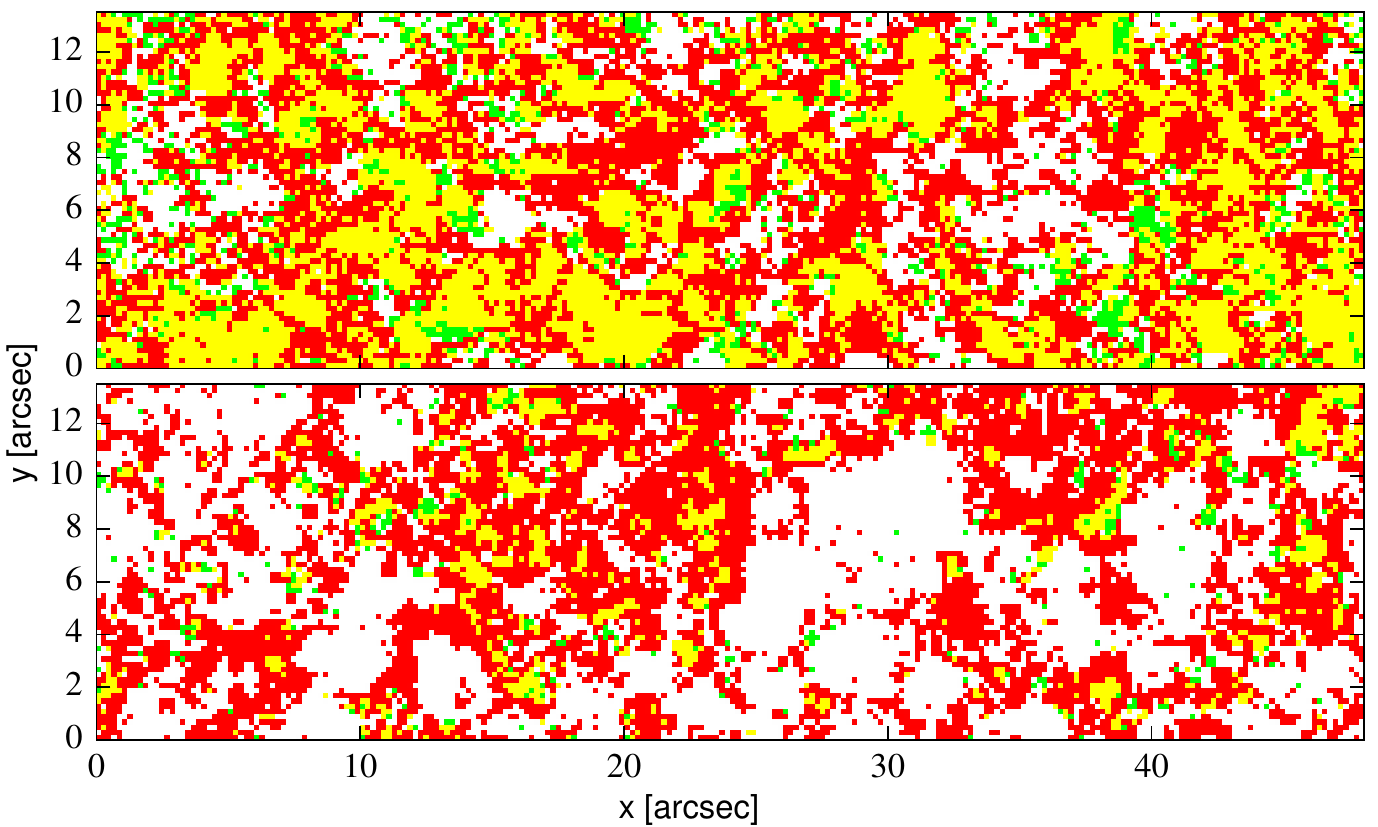}{Signal-to-noise ratio comparison between the GRIS (top panel) and SOT/SP (bottom panel) quiet-Sun scans. Red, green, and yellow pixels indicate where the CP, LP, or both signals are above the 3$\sigma$ level, respectively.}

Although both the GRIS and the SOT/SP scan describe a very quiet solar area, the SOT/SP scan shows slightly more Stokes $V$ network-like fields than the GRIS scan, which is indicative of a slightly higher magnetic activity. Despite this, the percentage of profiles above a certain noise threshold is significantly higher for the GRIS scan. \Fig{Noise.pdf} compares the area covered by pixels with signal levels above 3$\sigma$ (with $\sigma$ being the average root mean square value of the Stokes $Q,U,V$ spectra outside the spectral lines)  for the GRIS observations in the \fea{} line (top panel) and the SOT/SP scan in the \feahin{} line (bottom panel). Red regions highlight areas where only the CP signal is above the 3$\sigma$ level, green regions indicate where only the LP signal ($Q$ or $U$)  is above the 3$\sigma$ level, and yellow regions are for pixels where both LP and CP are above 3$\sigma$.  The white areas indicate regions where all Stokes parameters are below the 3$\sigma$ level. For the GRIS scan 79.7\% of the profiles are above 3$\sigma$ in at least one of the polarized Stokes parameters, whereas this percentage is 51.4\% for the SOT/SP profiles. 

Another remarkable difference is the extent of green and yellow regions, that is, the regions with $\mathrm{LP} \ge 3\sigma$, which is more than four times larger for the GRIS data than for the SOT/SP data (39.7\% vs. 9.8\%).
The transverse component of the magnetic field, that is, the field parallel to the solar surface for our scan recorded at disk center, produces a measurable signal in these regions. This is  a necessary prerequisite for the unambiguous computation of the magnetic field strength and orientation in quiet regions from inversions. The high magnetic sensitivity of the \fea{} line uncovers these signals, which remain hidden in the SOT/SP scans. The capability of detecting CP signals is only a factor of $\approx$1.5 higher for GRIS.

It should be mentioned that \citet{bellotrubio:12a} were able to increase the percentage of LP profiles in a Hinode fixed-slit observation up to 27\% (4.5$\sigma$) by integrating over 67~seconds at the expense of reduced spatial resolution because of solar evolution. It cannot be ruled out either that the absence of voids without visible Stokes signals in the GRIS data is partly caused by the smearing of stronger signals by scattered light, which is very weak in Hinode/SP measurements.

\Tab{tab:sigma} presents these statistics for two additional sigma levels. The increased magnetic sensitivity of GRIS compared to SOT/SP is reflected with approximately the same factor for 4$\sigma$ and 5$\sigma$ as well. It is worth mentioning that almost 50\% of the GRIS map is above the 5$\sigma$ threshold in at least one of the polarization profiles.

\begin{table}
  \caption[]{Percentage of linear (LP) and circular (CP) polarization profiles above a certain $\sigma$-threshold for GRIS and SOT/SP data sampled at 0\farcs20.}
  \label{tab:sigma}
  \begin{tabular}{c|cccc|cccc}        
    \hline\hline 
    $\sigma$- & \multicolumn{2}{c}{GRIS [\%]} & LP & LP & \multicolumn{2}{c}{SOT/SP [\%]} & LP & LP \\
   level & & & and & or & & & and & or \\
            & LP       &  CP   & CP   &  CP &   LP   &  CP& CP &  CP  \\ \hline 
  3$\sigma$ & 39.7 &  73.0 &  33.1 &  79.7 &   9.8 &  49.3 &   7.7 &  51.4 \\
  4$\sigma$ & 18.4 &  57.0 &  13.9 &  61.5 &   4.2 &  37.1 &   3.1 &  38.2 \\
  5$\sigma$ &  9.2 &  44.2 &   6.2 &  47.2 &   2.1 &  28.5 &   1.5 &  29.1 \\
    \hline
  \end{tabular}
\end{table}

\section{Complexity of profiles\label{complexity}}

\subsection{Multilobed Stokes $V$ profiles\label{lobecalc}}

Complex magnetic and velocity structures within the solar atmosphere can produce complex Stokes $V$ profiles. Velocity gradients along the line of sight produce asymmetric profiles, in extreme cases even with multiple lobes, which are often indistinguishable from the profiles produced by multiple atmospheric components within a resolution element (i.e., unresolved fine structure). Synthesized spectra from MHD simulations demonstrate that the increase in spatial resolution generally leads to a further increase in the complexity of the Stokes profiles, particularly in quiet-Sun regions. In the absence of instrumental degradation, the extreme conditions in small-scale features are not smeared out anymore. However, when the spatial resolution is sufficient to resolve the solar features and in the absence of complex line-of-sight velocity stratifications, the Stokes profiles should become simpler again.

Spectral lines with a response over a broad height range are particularly likely to produce highly complex profiles. The analysis of these profiles, which is usually performed using Stokes inversion techniques, requires the use of height-dependent atmospheres with many node points in height and therefore many free parameters. This is a special obstacle for the interpretation of weak Stokes signals because the information necessary to constrain the many free parameters is not available. 

High spatial resolution in combination with a narrow height range for the formation of the spectral line should therefore simplify the Stokes profiles and consequently their analysis. The GRIS data in the \fei{} 1.56\,$\mu$m range fulfill these requirements. The response functions (RFs) of these lines usually show relatively narrow peaks in deep photospheric layers \citep[for a detailed RF calculation see, e.g., ][]{borrero:16a}. In this section we analyze the complexity of the profiles by considering  the number of  Stokes $V$ lobes and by determining the amplitude and area asymmetries of the Stokes $V$ profiles.


We computed the number of lobes $l$ in the Stokes $V$ signal using the following scheme: Before the analysis, a three-pixel median filter was applied to the Stokes $V$ profiles. In the Stokes $V$ profile plots we then drew a horizontal line at the highest value of $V$. This line was gradually moved down toward lower $V$ values. If at least two consecutive points lay above this line and if the line was still above the primary threshold $n_p\sigma$ (in our case set to 3$\times$ the noise level, i.e., $n_p=3$), we found the primary lobe. When this primary lobe was identified, the detection threshold for the secondary lobe $n_s\sigma$ was set to a value lower than or equal to the primary threshold ($n_s\in[3,\,2,\,1.5,\,1]$). The horizontal line was then moved downward until it reached this secondary threshold $n_s\sigma$. At every step of this downward movement, the contiguous Stokes $V$ regions with at least two points lying above this horizontal line were counted. If such a contiguous region did not overlap with a region from the previous step, it was counted as a new lobe, otherwise the existing lobe was extended. This procedure was repeated for the part of the Stokes profile of the opposite sign, if present, with the detection threshold set to $n_s\sigma$. We note that this computation does not count a shoulder in a Stokes $V$ profile as an additional lobe, and therefore it may underestimate the number of complex profiles.

This lobe-counting method was applied to the GRIS and the Hinode SOT/SP data set, both resampled to the same spatial resolution (0\farcs20$/$pixel, see \sect{grisobs}). Stokes $V$ profiles with $l=1$ therefore represent single-lobed profiles, $l=2$ are two-lobed, roughly antisymmetric profiles (referred to as ``normal'' profiles, i.e., exhibiting the standard shape with a blue and a positive lobe of opposite sign), and $l\ge3$ are complex, multilobed profiles. For a discussion of three-lobed profiles measured in the penumbra with GRIS we refer to \citet{franz:16a}.


\begin{table}
  \caption[]{Lobe statistics for the GRIS and the Hinode spectral lines.}  \label{tab:lobe}
\noindent\begin{tabularx}{\linewidth}{c|cc|c|ccccc}\hline\hline 
 \fei{} & \multicolumn{2}{c|}{criteria} &   \% of&\multicolumn{5}{c}{\% of $V$ with $l=\ldots$ lobes [\%]}\\
line & $n_p$ & $n_s$ & total &  1 & $+1/-1$ & 2 & 3 & 4 \\\hline 
\multirow{4}{*}{\rotatebox{90}{15648.5\,\AA{}}}
&3.0 & 3.0 &  66.4 &  30.9 & 64.8 & 65.0 &  4.0 &  0.1 \\
&\bf 3.0 &\bf 2.0 &\bf 71.3 &\bf 18.3 &\bf 71.0 &\bf 71.5 &\bf 9.8 &\bf 0.4 \\
& 3.0 & 1.5 &  78.7 &  19.7 & 61.0 & 62.3 & 16.6 &  1.4  \\
& 3.0 & 1.0 &  88.2 &  22.1 & 42.4 & 46.5 & 25.6 &  5.3  \\
 \hline
\multirow{4}{*}{\rotatebox{90}{6302.5\,\AA{}}}
&  3.0 & 3.0 &  46.3 &  49.3 & 45.4 & 48.5 &  2.3 &  0.0\\
&\bf 3.0 &\bf 2.0 &\bf 49.8 &\bf 32.5 &\bf 55.2 &\bf 60.5 &\bf 6.8 &\bf 0.3 \\
&  3.0 & 1.5 &  58.0 &  30.5 & 46.9 & 52.9 & 14.8 &  1.8  \\
&  3.0 & 1.0 &  77.8 &  33.7 & 19.8 & 29.4 & 23.6 & 10.2  \\
\hline
\end{tabularx}
\end{table}

\Tab{tab:lobe} summarizes the result of this lobe analysis for the \fea{} and the \feahin{} lines for a detection threshold of $3\sigma$ for the primary (i.e., strongest) lobe and for different detection thresholds for the secondary lobe ($n_s\sigma$, with $n_s \in [3,\,2,\,1.5,\,1]$). The second column specifies the $\sigma$-thresholds used to detect the primary lobe ($n_p\sigma$) and the secondary lobe ($n_s\sigma$). The third column gives the percentage of pixels with at least one lobe ($l\ge1$), the remaining columns specify the relative percentage for $l=1\,\ldots\,4$ lobed profiles, regardless of their polarity, whereas the column labeled $+1/-1$ lists the percentage for the ``normal'' profiles with exactly one positive and one negative lobe. The boldface rows list the default $\sigma$-thresholds used later in this work for the analysis of the Stokes profiles. These thresholds were $n_p=3$ for the primary lobe and $n_s=2$ for the secondary lobe, with the additional requirement of two consecutive points lying above these thresholds. A larger $n_s$ means that we might be missing a number of weaker lobes, so that we tend to overestimate the number of normal or single-lobed profiles. However, the results are hardly affected by noise (particularly because we expect two neighboring pixels to lie above this threshold). As the threshold is lowered, a larger number of complex profiles is found, but at the cost of increasing influence of noise.

For the default $\sigma$-threshold settings ($n_p=3$, $n_s2$), 71\% of the $V$ profiles observed in \fea{} are simple, two-lobed profiles with one positive and one negative lobe (Col. $+1/-1$ in \tab{tab:lobe}). The column labeled $l=2$ contains all these normal profiles and the two-lobed profiles where both lobes have the same polarity, making up only a very small fraction of all two-lobed profiles. Single-lobed profiles contribute $\approx$19\% and three-lobed profiles $\approx$8\%. Slightly more than 55\% of the Hinode \feahin{} line are of regular, two-lobed shape. Single-lobed ($l=1$) profiles are more abundant for the Hinode lines, more complex profiles are relatively rare for GRIS and Hinode profiles. With lower thresholds for the secondary lobe $n_s$, the number of complex ($l\ge3$) profiles increases (see \tab{tab:lobe}). This can be a result of weaker lobes now being above the threshold and therefore being counted as new lobes, but also of an increasing number of false lobe detections caused by photon noise.

The default threshold settings of $3\sigma$ for the primary lobe and $2\sigma$ for the secondary lobe, listed in boldface font in \tab{tab:lobe}, were selected as a compromise between maximizing the number of detected minor lobes on the one hand and on the other hand  keeping the number of false detections caused by noise low. However, it should be noted that the various threshold settings have only a minor influence on the result of these analyses.

We note that the number of single-lobed profiles differs significantly from those published by \citet{sainzdalda:12}, who found only $\approx$5\% of the measured quiet-Sun Hinode profiles at disk center to be single-lobed. The reason for this discrepancy is the different definitions used to detect these profiles. The analysis of \citet{sainzdalda:12} requires the reliable detection of single-lobed profiles alone, therefore defining a 4$\sigma$ minimum threshold for one lobe, and a 3$\sigma$ maximum threshold for the other lobe. Our analysis focuses on the reliable detection of normal, two-lobed profiles, which requires the thresholding described above.

\subsection{Stokes $V$ asymmetries}

Asymmetries between the blue and red lobes of the Stokes profiles are a good indicator for the presence of velocity and magnetic field gradients along the line-of-sight direction. They have been analyzed extensively since spectropolarimetric measurements became available. Recent examples for an application of the asymmetry analysis to quiet-Sun data sets are \citet{viticchie:11} for Hinode SOT/SP and \citet{martinezgonzalez:12a} for \sunrise{}/\imax{} observations. The most sensitive measures for these asymmetries in the magnetized atmosphere are the Stokes $V$ amplitude ($\delta a$) and area asymmetries ($\delta A$), defined as 
\begin{eqnarray}\label{eq:asym}
\delta a &=& (|a_b|-|a_r|)/(|a_b|+|a_r|) \mbox{,~and}\\
\delta A &=& (|A_b|-|A_r|)/(|A_b|+|A_r|),
\end{eqnarray}
where $a$ and $A$ correspond to the amplitude and the area of the blue (subscript $b$) and the red (subscript $r$) lobe of the Stokes $V$ profile, respectively. This definition was used in previous studies, for example, \citet{solanki:84,stenflo:87,sigwarth:99,khomenko:03}. 

\colfig{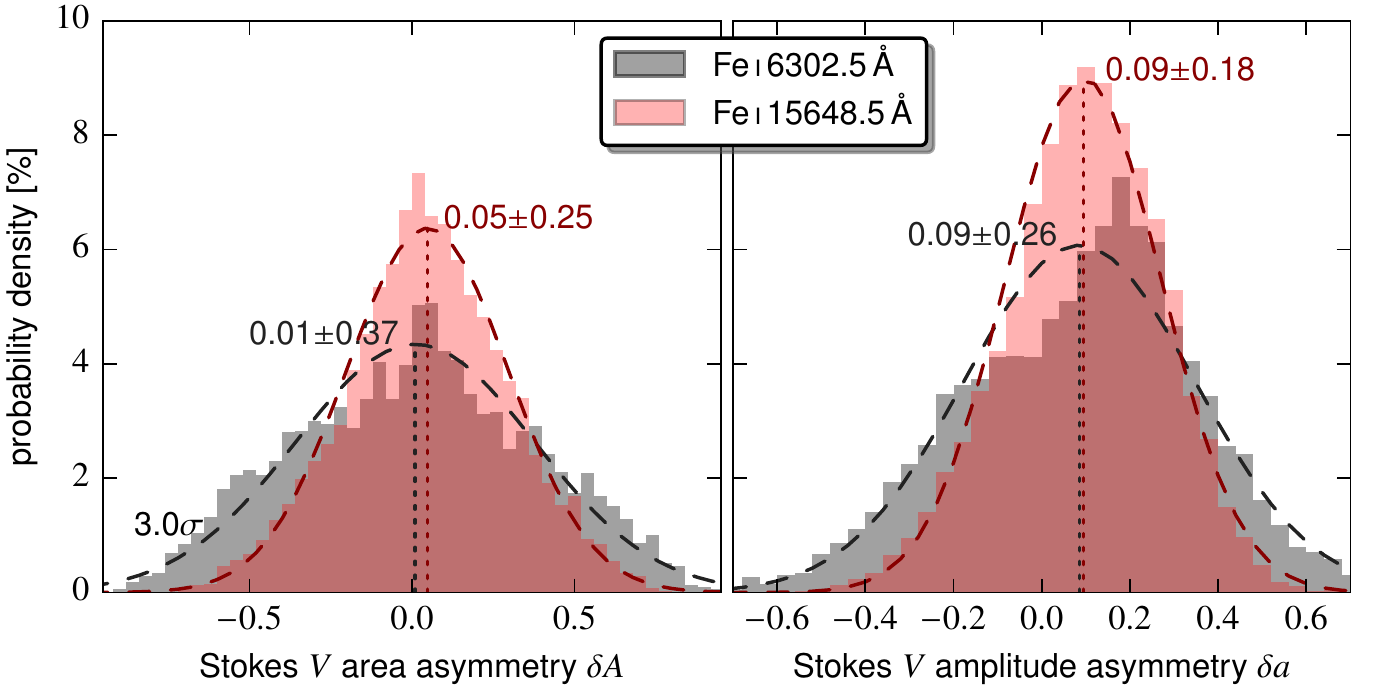}{Stokes $V$ area (left) and amplitude (right) asymmetry for two-lobed profiles in the GRIS \fea{} (red) and the SOT/SP \feahin{} (gray) lines. The dashed lines represent the fitted normal distributions, the dotted lines show the mean values. The numbers indicate the mean value and the standard deviation of the fitted normal distributions.}

\Fig{asym_histo_sp.pdf} compares the amplitude and the area asymmetries of the GRIS scan with the Hinode SOT/SP scan for all two-lobed Stokes $V$ profiles with a positive and a negative lobe\footnote{The single-lobed profiles were excluded from the asymmetry analysis since they would result in $\delta A$, $\delta a$ values of $\pm1$.}. 50.6\% of the $V$ profiles in the GRIS scan and 27.5\% in the SOT/SP scan are such normal profiles. The mean values for both asymmetries is positive for both data sets, in agreement with previous studies that used data from the Tenerife Infrared Polarimeter \citep{khomenko:03,martinezgonzalez:08b} or MHD simulations \citep{khomenko:05b}. According to \citet{solanki:88b} and \citet{solanki:93a}, the sign of the area asymmetry produced by a field strength or inclination gradient follows the equations
\begin{eqnarray}
\mbox{sign}(\delta A) &=& \mbox{sign}\left(-\frac{d|B|}{d\tau} \frac{dv}{d\tau}\right) \mbox{, and} \\
\mbox{sign}(\delta A) &=& \mbox{sign}\left(-\frac{d|\cos\gamma|}{d\tau} \frac{dv}{d\tau}\right),
\end{eqnarray}
with $\gamma$ being the magnetic field inclination to the line of sight, $\tau$ the optical depth, and $v$ the line-of-sight velocity, with positive velocities denoting downflows \citep[see also][]{solanki:93c}. The sign of the amplitude asymmetry depends on the details of the line-of-sight velocity gradient.


All asymmetries measured with GRIS have positive mean values. The area asymmetry of the GRIS data (red histogram in the left panel of \fig{asym_histo_sp.pdf}, $\delta A = 0.05\pm0.25$) has a lower mean value and slightly higher standard deviation than in \citet{khomenko:03}, who measured $\delta A$ of the same spectral line (\fea{}) at a spatial resolution of 1\arcsec{} and found $\delta A = 0.07\pm0.12$. The SOT/SP data has a mean value close zero ($\delta A=0.01$) but a significantly enhanced standard deviation (37\%, gray histogram in left panel of \fig{asym_histo_sp.pdf}).

The mean value of the amplitude asymmetry of the GRIS scan is with a value of 0.09 lower than the value obtained by \citet{khomenko:03} ($\delta a = 0.15$) and identical to the value for the SOT/SP scan ($\delta a = 0.09$). Similar to the area asymmetry, the amplitude asymmetry of GRIS also shows a significantly smaller standard deviation than SOT/SP (GRIS: 18\%, SOT/SP: 26\%).

The lower mean value and larger standard deviation of $\delta A$ in the high-resolution GRIS data than in the 1\arcsec{} resolution data analyzed by  \citet{khomenko:03} agree with the analysis by \citet{khomenko:05b} using MHD data, who showed that a higher spatial resolution decreases the mean value of the asymmetries and increases the standard deviation. In the extreme case of infinite spatial resolution, $\delta A$ values computed from MHD simulations tend to lie close to zero \citep[][and green histogram in the left panel of \fig{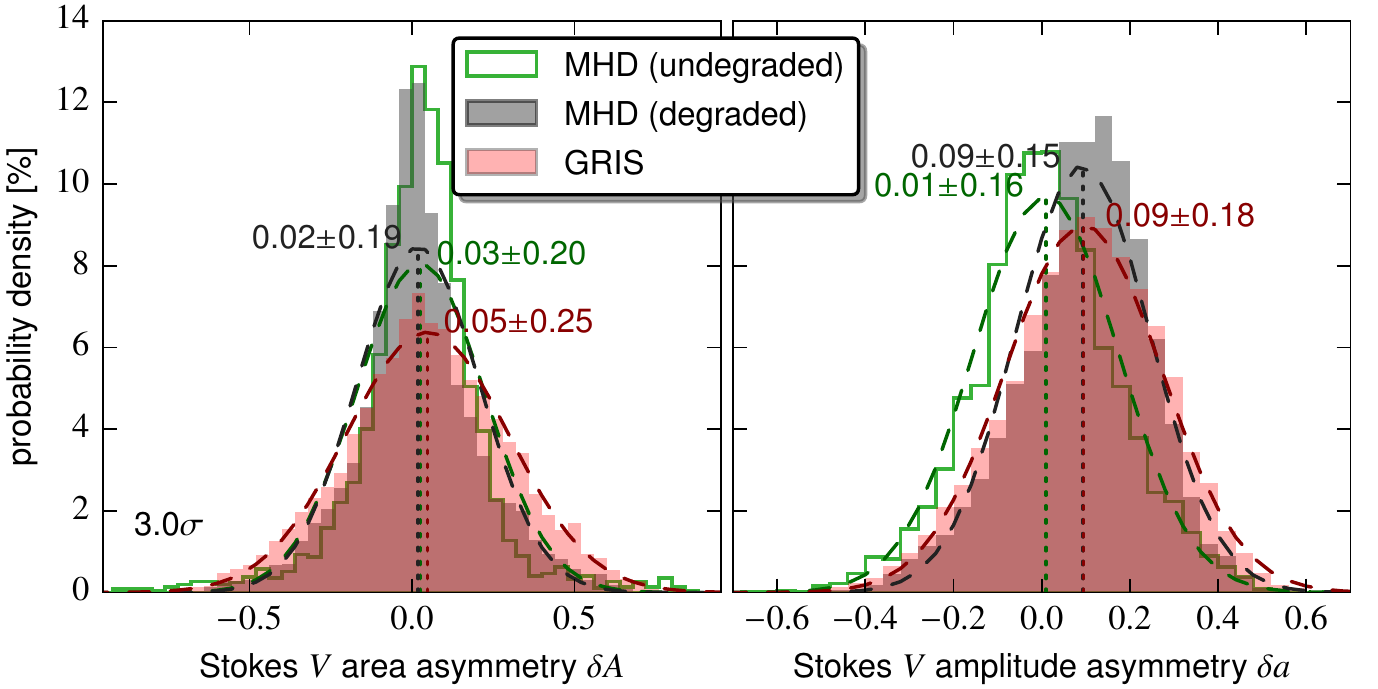}]{steiner:99, sheminova:04}. However, an increase of the mean value of $\delta A$ can also be a result of weak mean magnetic fields in the MHD simulation box.

\colfig{asym_histo_mhd.pdf}{Stokes $V$ area (left) and amplitude (right) asymmetry for two-lobed profiles in the \fea{} line for GRIS data (red), the degraded MHD data (gray), and the undegraded MHD data (green). The dashed lines show the fitted normal distributions, the dotted lines the mean values.}

\Fig{asym_histo_mhd.pdf} compares the asymmetries of the GRIS scan (red histogram, same as in \fig{asym_histo_sp.pdf}) with those from the undegraded (green histogram) and the spatially and spectrally degraded MHD data (gray histogram) described in \sect{mhddata}. The  undegraded Stokes $V$ profiles from the MHD cube do indeed display asymmetry values centered closely on zero, while  the agreement between the degraded MHD data and the GRIS data of the probability density distributions for both asymmetries, especially for $\delta a$, is rather good. We therefore conclude that the mean Stokes $V$ asymmetry in our data set that is lower than the 1\arcsec{} resolution data analyzed by  \citet{khomenko:03} is in fact a result of the higher spatial resolution.

The main difference in the asymmetries of the data sets used here (i.e., GRIS, SOT/SP, and MHD) is the significantly higher standard deviation of the SOT/SP scans in area and amplitude asymmetry. A possible explanation is the broader height range over which the \feahin{} line is formed. As a consequence, velocity and magnetic field gradients in height leave stronger imprints in the Stokes spectra than for lines with narrower formation height ranges, such as the \fea{} line. The combination of the low standard deviation of the asymmetries and the high percentage of normal Stokes $V$ profiles  makes the MLR technique applicable for a large portion of pixels in the GRIS scan.

\section{Magnetic line ratios\label{mlr}}

The MLR technique allows conclusions about the magnetic field strength drawn directly from the Stokes $V$ profiles of two spectral lines. Introduced by \citet{stenflo:73}, this technique circumvents some inversion problems of the radiative transfer equation especially for profiles with a low signal-to-noise ratio. This is the standard method for deriving the magnetic field and other atmospheric parameters from spectropolarimetric measurements. The idea behind the MLR method is that in the regime of incomplete Zeeman splitting (weak field regime), the amplitudes of the Stokes $V$ profiles (taken here as the larger of the blue and red amplitudes) scale with the magnetic flux (i.e., the magnetic field strength times the fill fraction of the magnetic structure embedded in an unmagnetized environment). It can be shown that if two spectral lines are formed under identical atmospheric conditions and have the same sensitivity to thermodynamics but a different Land\'e factor, the ratio of the amplitude of these two lines directly depends on the intrinsic magnetic field strength alone, and the dependence on the fill fraction is removed \citep{stenflo:73}.

The MLR technique, also called the Stokes $V$ amplitude ratio technique, works reliably under the following assumptions \citep[see also][]{steiner:12}: (1) The two spectral lines must have the same formation process, meaning that they need a very similar excitation potential of the lower level (e.g., the two lines belong to the same multiplet), oscillator strength, and wavelength. This ensures that the lines are formed at roughly the same heights, thus sampling almost the same atmospheric parameters, and that they are equally sensitive to temperature. (2) The resolution element producing the Stokes $V$ profile contains a magnetic field of only a single polarity. (3) The line-of-sight component of the field dominates (i.e., small $Q,U$ profiles), and (4) the magnetic field strength is below the value that leads to complete splitting in both spectral lines. For complete splitting, the amplitude ratio becomes constant and delivers only a lower limit to the field strength, namely the field strength at which the splitting just starts to be complete. 

Here we apply the MLR technique to the \fea{} and \feb{} lines. Unfortunately, these lines do not quite fulfill the requirement of identical line formation, but the formation height is sufficiently similar to obtain meaningful results. This is shown below in this chapter by applying the MLR technique to data from MHD simulations, where the known magnetic field can be used to test and validate this technique in a simple manner. The MLR for the \fea{} and \feb{} lines is defined as follows \citep[see also][]{solanki:92b}:
\begin{equation}\label{eq:mlr}
\mbox{MLR} = \frac{\mgeff{}(15652) \mvmax{}(15648)}{g(15648) \mvmax{}(15652)},
\end{equation}
with $g$ (and \tgeff{}) being the (effective) Land\'e factors, and \tvmax{} the Stokes $V$ amplitudes (maximum value of blue and red lobe). In a thorough analysis based on MHD simulations, \citet{khomenko:07} demonstrated the applicability of the MLR method to this line pair, especially its power to detect kilo-Gauss fields in internetwork regions. Unlike the Hinode SOT/SP line pair (\febhin{} and \feahin{}), the RFs for the infrared line pair are very similar and are moreover concentrated in a relatively narrow formation range. This, together with the high magnetic sensitivity of the \fea{} line, means that this line pair even outperforms the so far most successfully used pair for MLR-based magnetic field analyses, the \fei{} line pair at 5247/5250\,\AA{}, in particular for comparatively weak fields.

\colfig{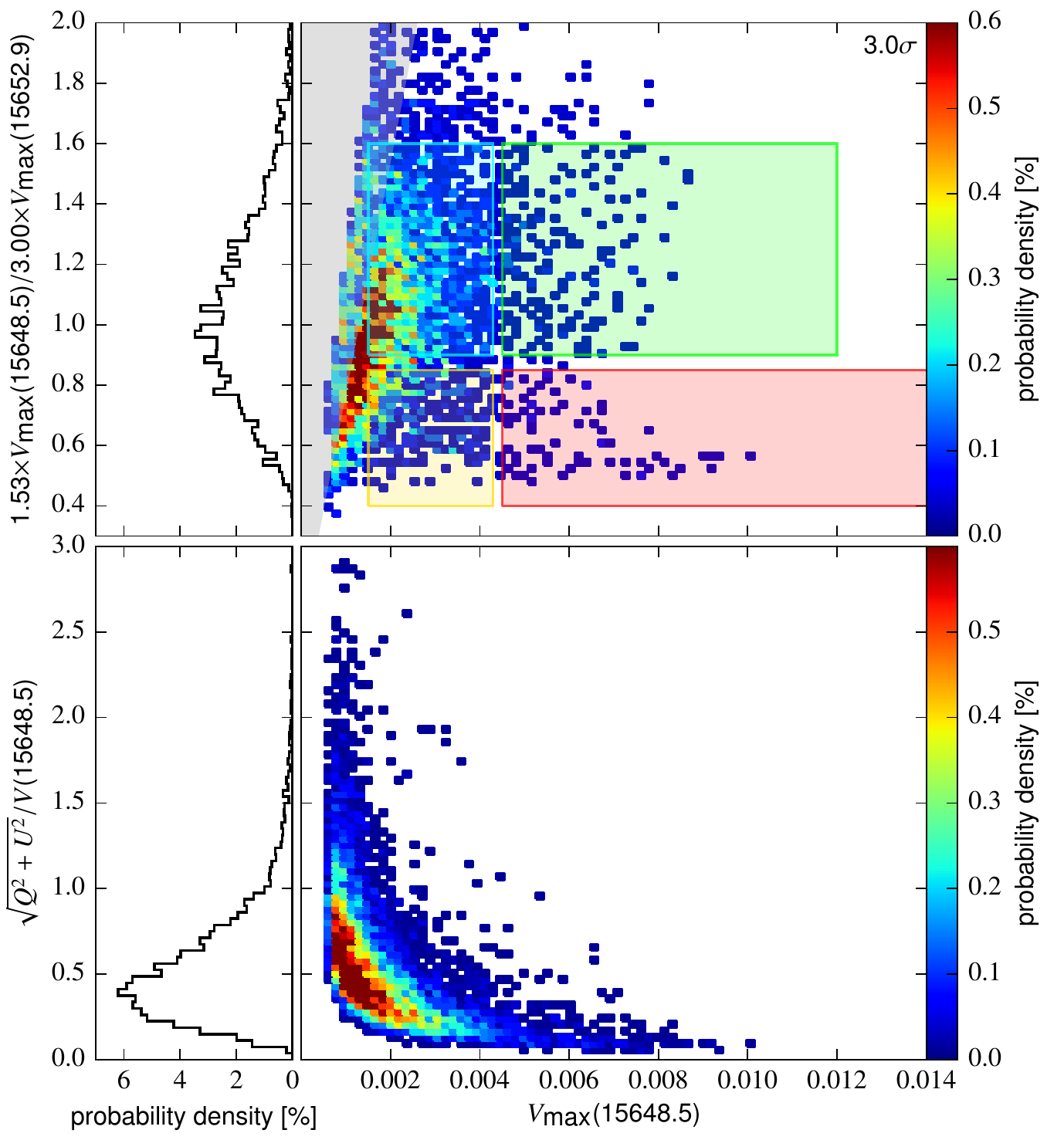}{Top panel: Magnetic line ratio (MLR) of the GRIS data set as a function of the Stokes $V$ amplitude \tvmax{} in \fea{}. The two-dimensional histogram takes only two-lobed profiles with small asymmetries ($\delta a$ and $\delta A$) into account. The left panel shows the histogram integrated over the $x$-axis of the scatter plot. The color-coded boxes mark regions discussed in the text, the gray wedge indicates the 3$\sigma$ threshold applied to compute the number of lobes. Bottom panel: same as above, but for the LP$/$CP ratio in the \fea{} line (see \sect{lp2cp}).}

\Fig{MLR-LP2V_gris.pdf} displays the MLR for the \fea{} and \feb{} lines obtained from our GRIS scan for all pixels where the Stokes $V$ profile in the \fea{} line is two-lobed (one positive and one negative lobe). In the right panel the probability density of the MLR versus \tvmax{} for the \fea{} line is plotted as a two-dimensional histogram. The left panel shows the same histogram integrated over the $x$-axis. To increase the reliability of the MLR analysis, we only used two-lobed profiles with small area or amplitude asymmetries ($|\delta a|, |\delta A|\le0.4$) to avoid pixels with strong velocity and possibly also magnetic field gradients. This restriction minimizes the effect that the two lines do not sample the exact same height layer. For the GRIS data 43.7\% of the profiles survive this thresholding.

We also tested an alternative definition for the MLR, in which the Stokes $V$ profiles are divided by the first derivative of the Stokes $I$ profile before applying \eq{eq:mlr}. This division by $dI/d\lambda$ should eliminate most of the non-magnetic effects on the MLR. The results of the MLR analysis using this alternative definition did not differ from the original definition, but they introduced a larger scatter in the MLR distribution. We therefore refrained from using this alternative definition.

The two-dimensional histogram in \fig{MLR-LP2V_gris.pdf} shows various distinct regions: the gray shaded wedge on the left side indicates the 3$\sigma$ threshold applied to the computation of the number of lobes (see \sect{lobecalc}). The blue and green boxes identify high MLR values around 1.2 for weak and strong \tvmax{} values, respectively. According to \citet{solanki:92b}, such high MLR values are indicative of fields of up to a few hundred Gauss, while MLR values around 0.6 point to the presence of kilo-Gauss fields within the observed pixel. These regions are shown by the yellow (low \tvmax{}) and red (high \tvmax{}) boxes. The latter contains a small but distinct population with large Stokes $V$ amplitudes and with an MLR around 0.6.

\colfig{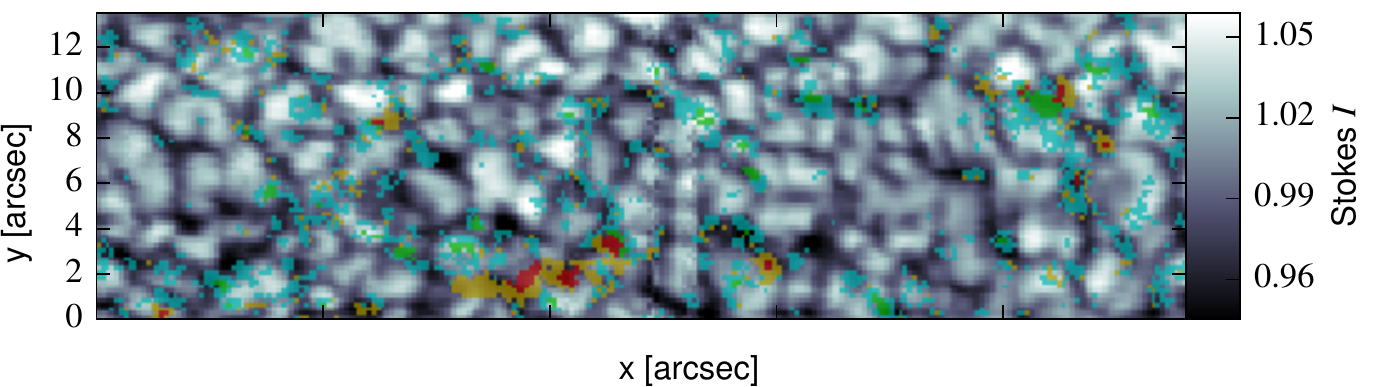}{Continuum intensity map from the GRIS scan (same as top panel in \fig{Stokes-gris.pdf}) with color overlays indicating the positions of pixels within the colored boxes in \fig{MLR-LP2V_gris.pdf}, marking specific ranges for \tvmax{} and MLR.}

The locations of the points within these colored boxes are marked with the same color coding in the continuum intensity map shown in \fig{Stokes-gris_I.pdf}. It is noticeable that the red regions, concentrated in intergranular lanes, are surrounded by yellow regions. The locations of the blue and green regions do not show a clear pattern and are equally distributed over granules and intergranular lanes.

Examples of Stokes $I$ and $V$ profiles for a low and a high MLR are displayed in \figs{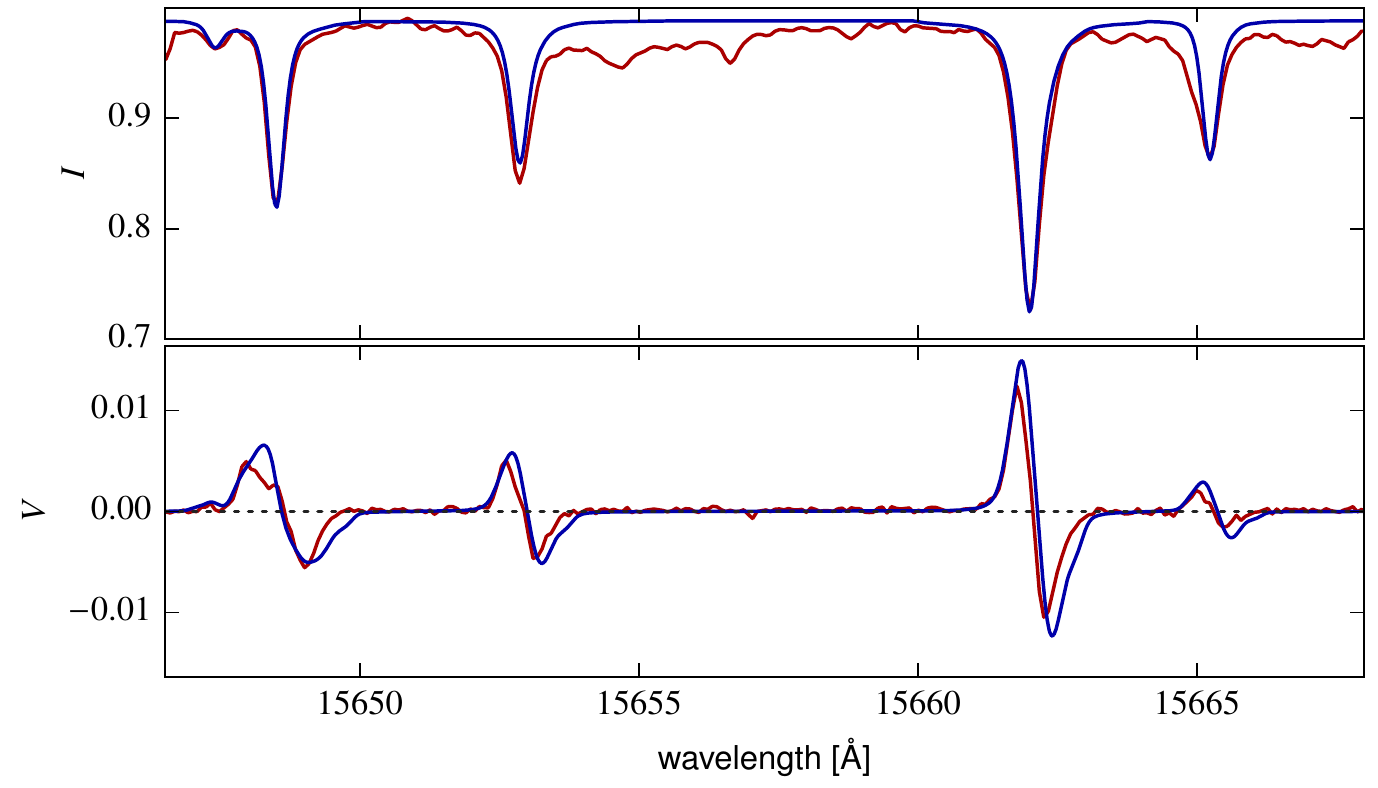} and \ref{prof_gris_mhd_hG.pdf}, respectively. The red profile shows the GRIS measurement, the blue line a synthetic profile from an MHD cube, discussed in the next section. The wavelength in this plot covers the four main \fei{} lines in the GRIS spectral range, the MLR technique was applied to the two left lines (\fea{} and \feb{}).

\colfig{prof_gris_mhd_kG.pdf}{Stokes $I$ and $V$ profile from the GRIS scan (red, indicated with a triangle in \fig{Stokes-gris.pdf}) and the MHD run (blue, indicated with a triangle in \fig{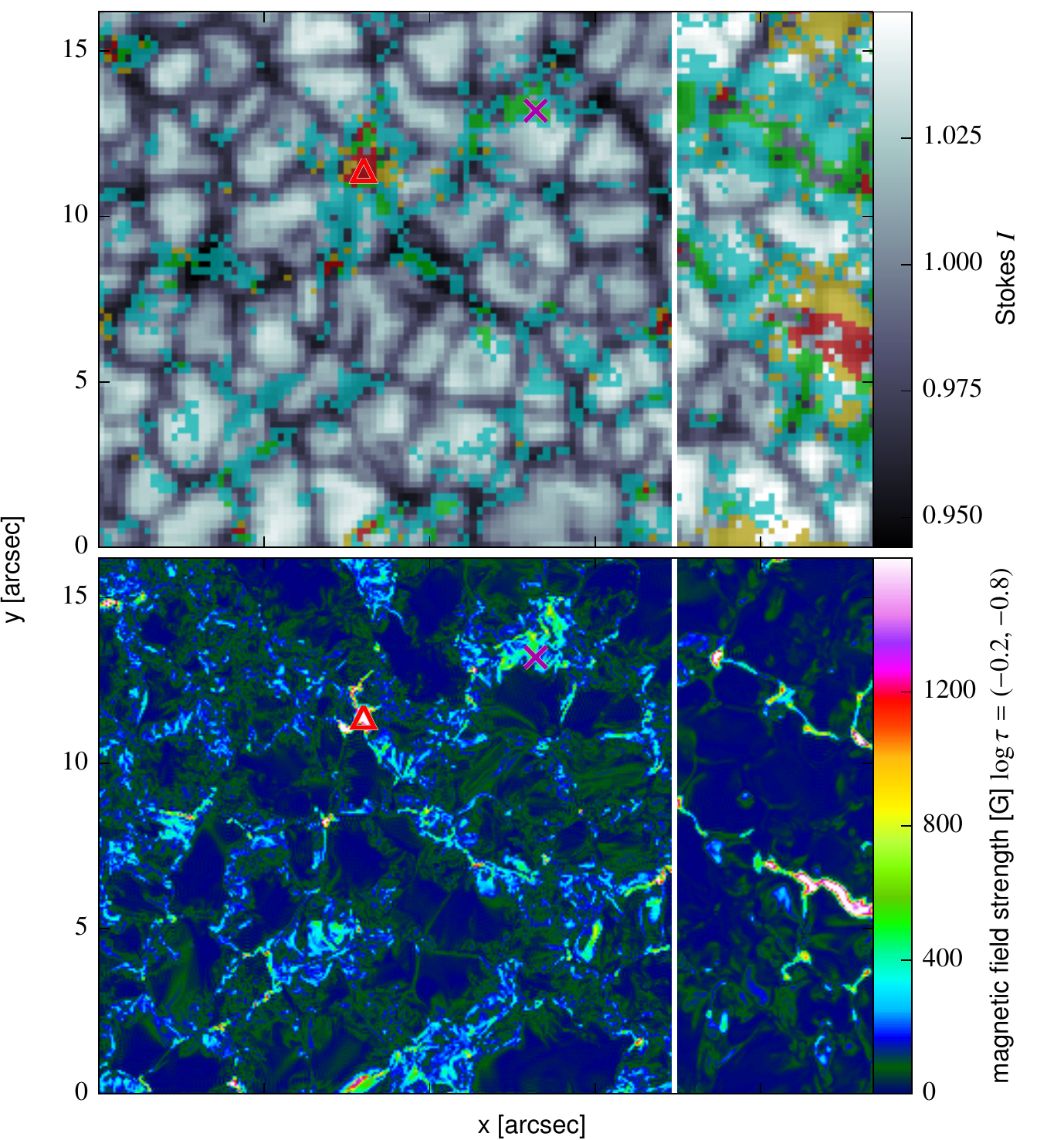}) with MLR=0.6 taken from the red boxes in \figs{MLR-LP2V_gris.pdf} and \ref{MLR-LP2V_mhdqs.pdf}. The MHD profile corresponds to a pixel with $B\approx2.0$\,kG averaged over $\log\tau=(-0.2,-0.8)$.}

\colfig{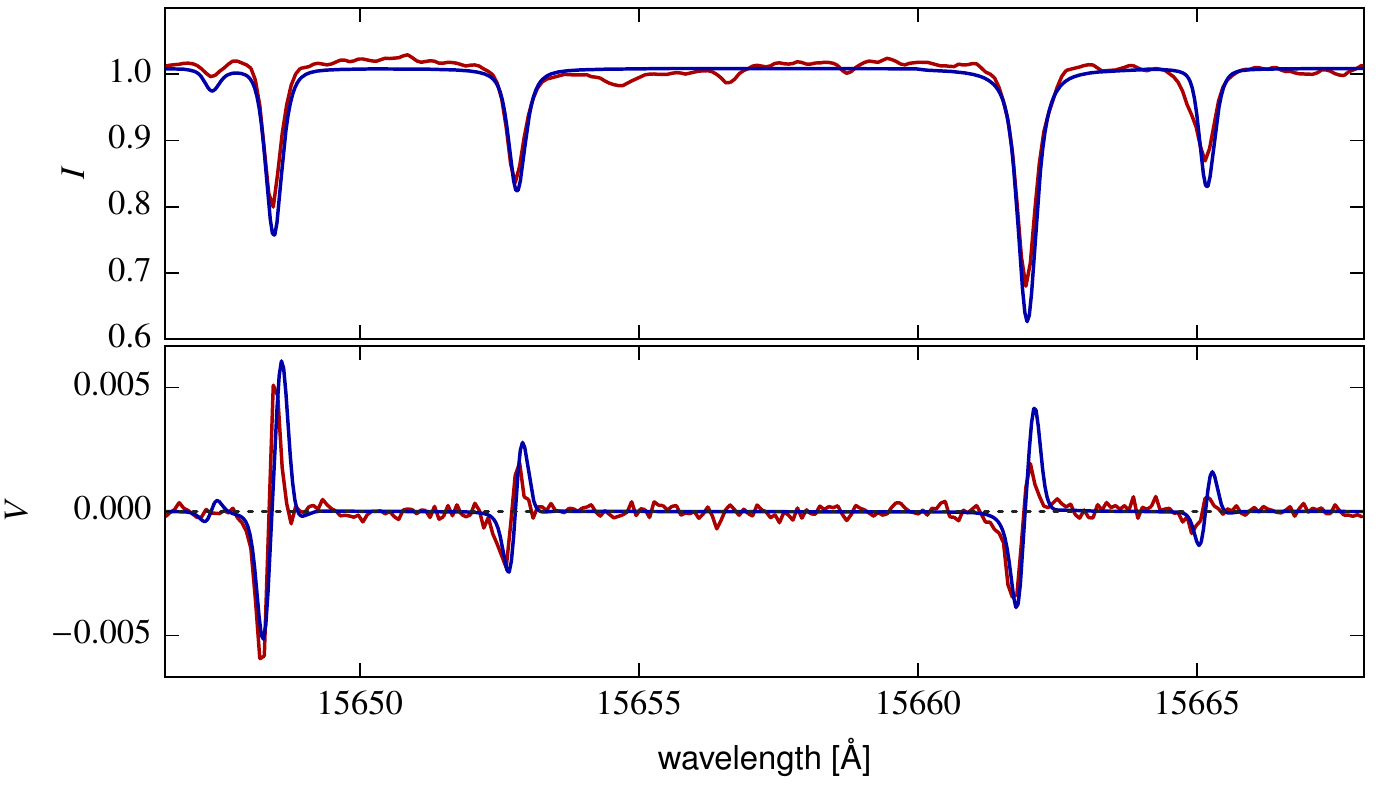}{Same as \fig{prof_gris_mhd_kG.pdf}, but for a profile with MLR=1.1 taken from the green box in \figs{MLR-LP2V_gris.pdf} and \ref{MLR-LP2V_mhdqs.pdf}. A cross shows the position of the profiles in the corresponding maps in \figs{Stokes-gris.pdf} and \ref{Stokes-mhdqs.pdf}. The MHD profile corresponds to a pixel with $B\approx200$\,G averaged over $\log\tau=(-0.2,-0.8)$.}

\subsection{LP$/$CP ratio \label{lp2cp}}

The ratio between the linear and the circular polarization (LP$/$CP ratio) carries information about the magnetic field inclination $\gamma$ with respect to the line of sight. We define the LP$/$CP ratio according to \citet{solanki:92b} as $\sqrt{\smash[b]{Q^2_{\mbox{\tiny{max}}}+U^2_{\mbox{\tiny{max}}}}}/V_{\mbox{\tiny{max}}}$, with the subscript 'max' defining the maximum absolute value for the corresponding Stokes profile (after applying a median filter over three wavelength pixels) of the spectral line. \citet{solanki:92b} demonstrated that this ratio only depends on the inclination angle $\gamma$ of the field with respect to the line-of-sight direction for magnetic field strengths above a certain threshold. This threshold depends on the magnetic sensitivity of the line and is with $\approx$1000\,G for the \fea{} line in quiet-Sun areas lower than for most other spectral lines. Below this field strength, the LP$/$CP ratio depends not only on $\gamma$, but also linearly on the magnetic field strength. We note that since the observed field of view is at disk center ($\mu=1$), no transformation of the inferred inclination from the line of sight into the local solar frame of reference is needed.

The LP$/$CP ratio for the GRIS data is plotted in the bottom panel of \fig{MLR-LP2V_gris.pdf} for all normal Stokes profiles, that is, where Stokes $V$ has only one positive and one negative lobe. The shape of the curve is dominated by the $1/\mvmax{}$ dependence, defined by the applied thresholds of $3\sigma$ and $2\sigma$ for the primary and secondary lobe, respectively. Stokes profiles with high \tvmax{} values clearly populate the region with low LP$/$CP ratios below 0.3, clearly correlated with an inclination angle of $\gamma\le20^\circ$. Almost all of these profiles originate from patches with field strengths $\ge$1\,kG. For low \tvmax{} values, the LP$/$CP ratio is distributed over a wide range; this is suggestive of the absence of a preferred inclination.

\subsection{Comparison to MHD simulations\label{mlrmhd}}

The MLR technique usually relies on determining a calibration curve, which establishes a relation between the magnetic field in standardized atmospheres and the MLR \citep[e.g.,][]{solanki:92b}. The MHD simulations described in \sect{mhddata} allow us to go a step further. By synthesizing the line profiles and the subsequent degradation in the spatial and spectral domain, the MLR can be computed and compared to the one determined from the GRIS data. The advantage of using the MLR technique on MHD cubes is of course the knowledge of the atmospheric conditions in every pixel of the map, in particular the magnetic field strength. In a scatter plot of MLRs from MHD data, similar to \fig{MLR-LP2V_gris.pdf}, the different regions should therefore be discernible by their magnetic field strength.

\colfig{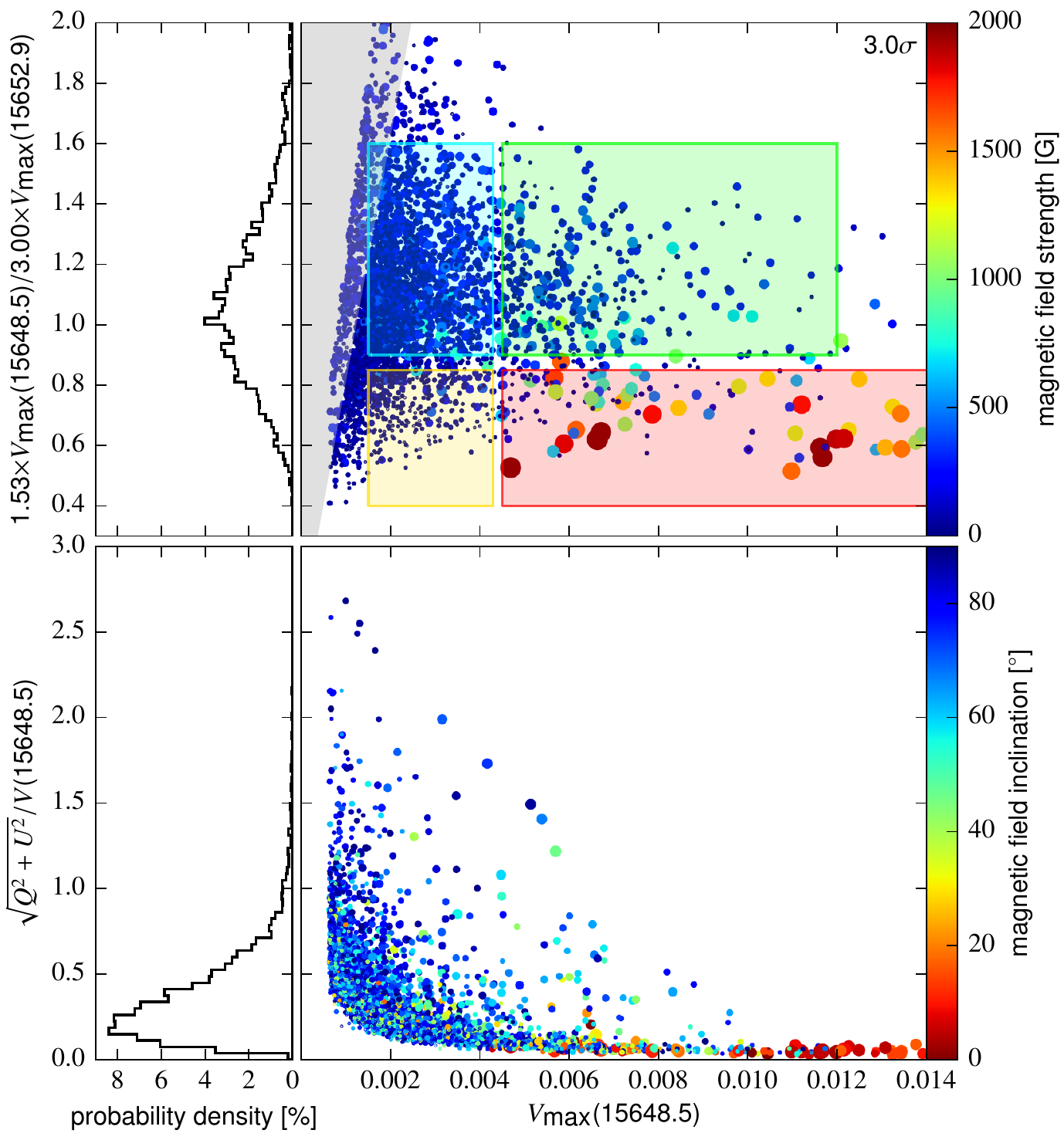}{Top panel: Scatter plot of the MLR computed from the spatially and spectrally degraded Stokes $V$ profiles of the MHD data as a function of \tvmax{} in \fea{}. The color coding and size of the symbols represent the magnetic field strength averaged over $\log\tau=(-0.2,-0.8)$. The two-dimensional histogram takes only two-lobed profiles with small asymmetries ($\delta a$, $\delta A\le0.4$) into account. The left panel shows a histogram of the line ratio to be comparable with the histogram in the left panel of \fig{MLR-LP2V_gris.pdf}. Bottom panel: same as above, but for the LP$/$CP ratio in the \fea{} line (see \sect{lp2cp}). Here the color coding represents the magnetic field inclination averaged over $\log\tau=(-0.2,-0.8)$.}

This scatter plot is presented in the top panel of \fig{MLR-LP2V_mhdqs.pdf}. It comprises data from the two MHD runs described in \sect{mhddata}, the small-scale dynamo run (MHD/SSD), and the MHD/\imax{} run, degraded to the spatial and spectral resolution of the GRIS scan. These two MHD runs were combined because we found that the MHD/SSD run did not contain as many low MLR values (corresponding to strong fields; see below) as present in the GRIS data. The $x$ and $y$ axes are identical to those in \fig{MLR-LP2V_gris.pdf}, but here the color coding and the size of the symbols represent the magnetic field strength from the  undegraded MHD cube, binned to the GRIS pixel size of 0\farcs20 and averaged over an optical depth range from $\log\tau=-0.2$ to $-0.8$, that is, the range over which the two lines collect a large part of their contribution \citep{borrero:16a}. \Fig{MLR-LP2V_mhdqs.pdf} reveals that the distribution of the points derived from the MHD data is very similar to the GRIS data: most of the pixels have  \tvmax{} values lower than 0.004, and the MLR displays a Gaussian distribution centered at unity.

The undegraded, binned cube for the scatter plots in \fig{MLR-LP2V_mhdqs.pdf} allowed us to study the reliability of the MLR and the LP$/$CP ratio technique to recover the true magnetic field configuration, regardless of the angular resolution of the telescope. A convolution of the magnetic field strength and inclination maps with the telescope PSF would smear out and therefore dilute the strong fields, originally confined to narrow regions in the intergranular lanes and their junctions, and therefore destroy the original magnetic field topology. The binning of the magnetic field strength and inclination maps, however, was necessary since the same binning was applied to the MLR and LP$/$CP ratios, computed from the PSF-degraded MHD Stokes profiles to make them representative of the GRIS observations. This binning has only a minor influence on the results presented here because it did not significantly lower the maximum field strengths or change the inclination in the strong field regions.

The largest part of the map in the GRIS and MHD data is covered by Stokes profiles with an MLR in the range of 0.9 to 1.4. The MHD data clearly reveal these regions to be populated by weak fields in the range of between a few Gauss and a few hundred Gauss.

The points with high magnetic field values of more than 1\,kG (green, yellow, and red bullets) are all located in the red box that shows the region with an MLR between 0.4 and 0.85 and $\mvmax{}\ge0.004$. This clearly demonstrates that a strong magnetic field is a sufficient condition for producing small MLRs. However, there are also blue points in the red box, indicating that field strengths in the ten to few hundred Gauss regime are also able to produce low MLR values. At first sight this suggests that the MLR technique does not correctly identify the kilo-Gauss Stokes profiles.

\colfig{Stokes-mhdqs.pdf}{Continuum intensity map of the MHD data (top, degraded to GRIS resolution) and magnetic field strength map (bottom, original MHD resolution) of the small-scale dynamo run (MHD/SSD, left three quarters of the map) and the MHD/\imax{} run (right quarter), separated by the white line. The color coding in the $I$ map (top) indicates the regions with specific ranges for \tvmax{} and MLR indicated in \fig{MLR-LP2V_mhdqs.pdf}. The triangle and cross show the positions of the Stokes profiles presented in \figs{prof_gris_mhd_kG.pdf} and \ref{prof_gris_mhd_hG.pdf}.} 

To gain insight into this problem, we consider the positions of the points in the colored boxes in the continuum image of \fig{Stokes-mhdqs.pdf}. The red and yellow points in the continuum map (top panel), denoting MLR values around 0.6, are without exception either overlapping kilo-Gauss magnetic fields (see bottom panel) or surrounding them like a halo. This effect can be explained by the smearing of the Stokes signal caused by the PSF of the telescope, applied to the MHD Stokes vector to simulate real observations. The imprint of the kilo-Gauss fields is visible in all Stokes profiles in an area where the wings of the PSF still contain significant power, in this case, a distance of approximately 1\,--\,2\arcsec{}.

The MLR technique is able to recover the kilo-Gauss fields from these diluted Stokes profiles. It is sufficient that a small fraction of the resolution element is covered by strong fields to lower the MLR to values around 0.6 because most of the $V$ profiles in regions with kilo-Gauss fields are strong and the MLR technique does not depend on the fill fraction. Conversely, this means that whenever we observe a Stokes profile with a low MLR value, it either corresponds directly to a kilo-Gauss feature or one must be in its vicinity.

The same color coding is also applied to the continuum map of the GRIS observation in \fig{Stokes-gris_I.pdf}. Similarly to the MHD maps in \fig{Stokes-mhdqs.pdf}, the red regions are surrounded by yellow regions. For the GRIS scan this suggests the presence of small-scale kilo-Gauss patches in the quiet Sun. These patches might well be smaller than the spatial resolution of the GRIS scan of 0\farcs40, surrounded by a halo of signals influenced by them (in analogy to the MHD maps). These halos are the result of straylight described by the PSF of the telescope. This is supported by the fact that the magnetic fields in the undegraded MHD maps (bottom panel of \fig{MLR-LP2V_mhdqs.pdf}) do not show such halos, but are confined to small patches mainly located in the junction of intergranular lanes. Since the areas around strong-field features are also populated by weaker field features, the effect of the straylight is to produce complex line profiles that are composed of the actual profile at a given location and the straylight from a stronger field region.

The bottom panel of \fig{MLR-LP2V_mhdqs.pdf} displays the LP$/$CP ratio as defined in \sect{lp2cp}. The shape of the distribution is similar to the one observed with GRIS (bottom panel of \fig{MLR-LP2V_gris.pdf}). The color coding, representing the line-of-sight inclination angle $\gamma$ (with the polarity information removed), visualizes that the low LP$/$CP ratios at high \tvmax{} values correspond to almost vertical magnetic fields. The remaining distribution shows a preference for more horizontal fields. 

\colfig{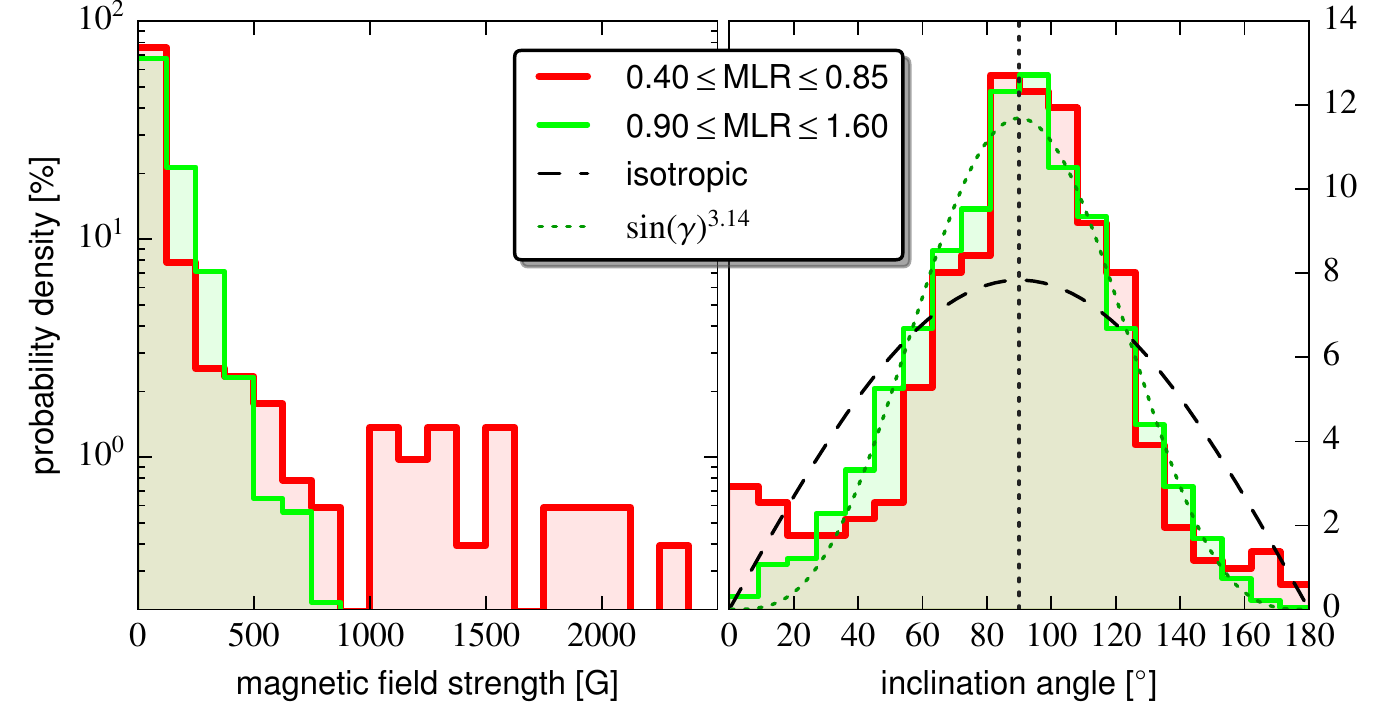}{Probability distribution for the magnetic field strength (left, logarithmic $y$-scale) and inclination (right, linear $y$-scale) from the MHD simulations (binned to the GRIS pixel size), in red for low MLR values (red and yellow boxes in \fig{MLR-LP2V_mhdqs.pdf}), and in green for high MLR values (blue and green boxes in \fig{MLR-LP2V_mhdqs.pdf}). The dashed line in the right panel indicates an isotropic distribution, the dotted line is a fit to the high MLR distribution (green) with the functional form $\sin(\gamma)^a$ with $a=3.67$.}

The good agreement between the MHD simulations and the GRIS scan in the asymmetry, the MLR and the LP$/$CP analyses suggests that the magnetic field in the simulation resembles the actual situation on the solar surface quite well. It is therefore worth looking at the distribution of the magnetic field strength and orientation in the MHD simulations. With a certain degree of caution, we can assume that these distributions are then also representative for the observed quiet-Sun area.

These distributions, computed from the degraded MHD Stokes profiles, are presented in \fig{b_inc_histo.pdf} for two different populations in the MLR scatter plot of \fig{MLR-LP2V_mhdqs.pdf}. The magnetic field strength and inclination were averaged over a $\log\tau$ range from $-0.2$ to $-0.8$ in the undegraded MHD cube and binned to the GRIS pixel size. The red histogram in both panels of \fig{b_inc_histo.pdf}  contains pixels with low MLR values (red and yellow boxes in \fig{MLR-LP2V_mhdqs.pdf}), the green histogram is for high MLR values (green and blue boxes in \fig{MLR-LP2V_mhdqs.pdf}). The magnetic field strength distribution clearly demonstrates that the strong fields (above 800\,G) are solely attributed to low MLR values (red histogram). In the inclination distribution (right panel), both histograms peak close to 90$^\circ$ (marked with the vertical, dotted line), with a slight asymmetry for the low MLR distribution toward lower inclination angles. This is indicative of a preferred polarity in the subfield selected for the comparison with the GRIS data. Compared to an isotropic distribution (dashed black line), both distributions show an overabundance of horizontal fields \citep[90$^\circ$ inclination angle, cf.][]{martinezgonzalez:16a}. The high MLR distribution can be well fit with the functional form $\sin(\gamma)^a$ with $a=3.14$, indicative of a redistribution of fields from an isotropic distribution toward a more horizontal one, leaving an underabundance of vertical fields. Only the red histogram, containing pixels with kilo-Gauss fields, exhibits an additional peak between 0$^\circ$ and 20$^\circ$. Apparently, these strong fields have the same polarity and are oriented vertically. We emphasize that the distributions presented in \fig{b_inc_histo.pdf} are a direct result of the MHD data and are therefore only indirectly related to the GRIS data.

\colfig{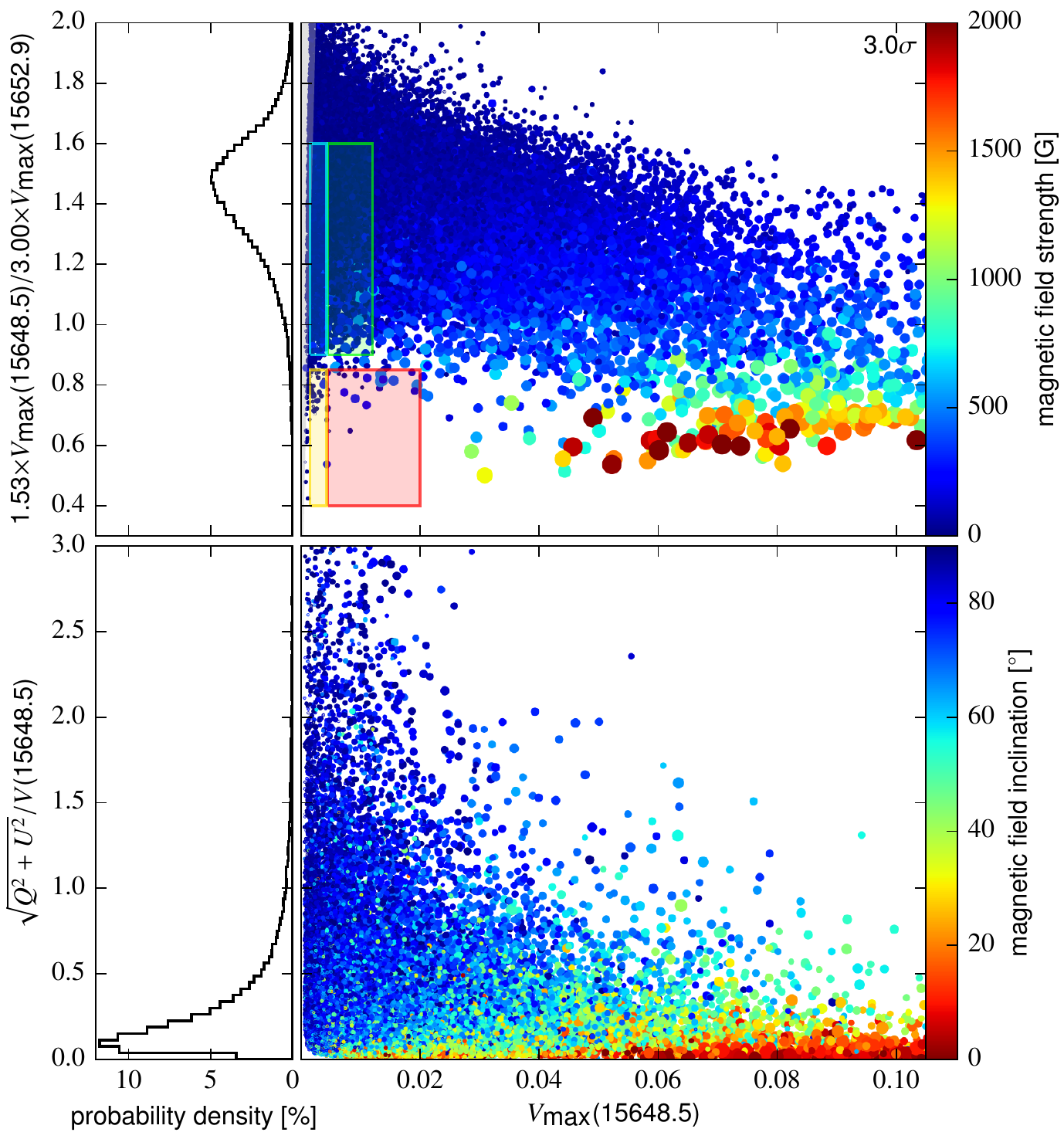}{MLR (top panel) and LP$/$CP (bottom panel)  as a function of \tvmax{} in \fea{}, computed from the undegraded MHD Stokes profiles. Similar to \fig{MLR-LP2V_mhdqs.pdf}, the color coding represents the magnetic field strength and the inclination angle, respectively, averaged over $\log\tau=(-0.2,-0.8)$ in the undegraded MHD cube. The colored boxes in the top panel mark the same MLR regions as in \figs{MLR-LP2V_gris.pdf} and \ref{MLR-LP2V_mhdqs.pdf}.}

The good agreement between the degraded MHD simulations and the observations motivate a plot similar to \fig{MLR-LP2V_mhdqs.pdf}, but now for the MHD data in its original resolution without any spatial or spectral degradation, to visualize the influence of the degradation on the MLR, the LP$/$CP ratio, and the \tvmax{} value. This plot is presented in  \fig{MLR-LP2V_mhdqs_orig.pdf} for all two- and three-lobed profiles with small asymmetries ($\delta a$, $\delta A\le0.4$). The most striking difference is the significantly higher values for \tvmax{}, demonstrating the strong influence of spatial and spectral degradation on the strength of the Stokes signals. Despite this huge change in signal strength, the ranges for the MLR and the LP$/$CP ratio remain in the same regime as for the degraded data, supporting the argument that these ratios are independent of resolution. Strong ($\ge$1\,kG) and vertical ($\le20^\circ$) fields clearly only populate the areas with low MLR and LP$/$CP ratios, respectively, but now at much higher \tvmax{} values. The weaker fields, in the few hundred Gauss range and below, produce Stokes profiles with MLR values higher than 0.9. The LP$/$CP plot (lower panel) indicates that most of the weaker fields, identified by low \tvmax{} values in the top panel, are predominantly horizontal. With increasing \tvmax{} values the fields become more vertical.

It is interesting to note that the low MLR values for the undegraded MHD data are exclusively produced by strong ($\ge1$\,kG) magnetic fields (see void area in red and yellow box in \fig{MLR-LP2V_mhdqs_orig.pdf}), in contrast to the degraded data, where weak field profiles can also result in low MLR values (see the blue symbols in the red and yellow boxes in \fig{MLR-LP2V_mhdqs.pdf}, top panel). \Fig{Stokes-mhdqs.pdf} reveals that these profiles are always located in the vicinity of strong field regions, supporting the argument that their low MLR values are a consequence of photons from these strong field regions, redistributed to pixels in their vicinity by the action of the PSF.

\section{Summary and conclusion\label{conclusion}}

We analyzed a very quiet solar region with very low magnetic activity, recorded at disk center with the GREGOR Infrared Spectrograph (GRIS) with an unprecedented combination of spatial resolution (0\farcs40) and polarimetric sensitivity (noise level $\sigma=2.2\times10^{-4}$ of the continuum intensity) in the \fei{} infrared lines around 1.56\,$\mu$m. About 80\% of the Stokes profiles in the map show polarization signals above a 3$\sigma$ threshold in at least one of the Stokes parameters, and 40\% of the linear polarization profiles exceed this level. This is a significant increase of the magnetic sensitivity compared to a deep magnetogram scan (12.8\,s exposure time per slit position) of Hinode SOT/SP, where these numbers are 51\% and 10\%, respectively.

The GRIS Stokes $V$ profiles show on average less scatter in area and amplitude asymmetries than the Hinode profiles. We attribute this fact to the narrower height of formation range of the infrared lines compared to the SOT/SP \fei{} line pair at 630\,nm. This minimizes the influence of gradients in the vertical velocity and the magnetic field on the Stokes profiles. In addition, the high Zeeman sensitivity of the IR lines means that larger velocity gradients are needed to produce a given asymmetry than for lines in the visible \citep{grossmanndoerth:89}, which also tends to result in smaller asymmetries. Stokes $V$ area and amplitude asymmetries agree well with small-scale dynamo MHD simulations \citep[run O16bM in ][]{rempel:14a}. We therefore conclude that the structure of the magnetic fields in the MHD/SSD run, in particular their vertical gradients, resemble the true conditions in the solar photosphere quite well.

The magnetic line ratio technique (MLR) reveals that the main part of the scanned region shows magnetic field strengths in a range from a few Gauss to a few hundred Gauss, indicated by high MLR values (0.9\,--\,1.4), and consistent with the MHD/SSD run. It also uncovers a few small-scale kilo-Gauss magnetic flux concentrations (MLR $=0.4$\,--\,0.85), which are underrepresented in the MHD/SSD run.  A comparison to MHD simulations, where we added the higher activity-level MHD/IMaX run to the MHD/SSD run, revealed that the signature of these flux concentrations extends into a halo of $\approx$1\,--\,2\arcsec{}, caused by the smearing of the signal because of the point spread function (PSF) of the telescope. The MHD simulations suggest that the weak field distribution shows an overabundance of magnetic fields parallel to the solar surface, whereas the strong magnetic fields are nearly vertical. This is also supported by the LP$/$CP ratio, indicating that these strong fields are nearly vertical to the solar surface, whereas the weaker fields do not show a clear preference in their inclinations.

The dilution of the signal will lead to erroneous results when computing the magnetic field value from the inversion of the Stokes profile, especially when used with simple, single-component atmospheric models. A correct inversion requires at least a two-component atmosphere composed of the magnetic flux concentration and the surrounding weakly magnetized area, which need to be mixed using a fill factor. In such a model this fill factor does not resemble the fraction of the pixel covered by the magnetic flux concentration, but takes the dilution of the signal caused by the PSF into account. However, introducing a fill factor as a free parameter can lead to ambiguous results regarding field strength and inclination \citep{orozco:07c}. The preferred approach for inverting these data is to take the signal dilution by the PSF self-consistently into account during the inversion process or to deconvolve the data with the PSF before the inversion. Such inversion methods do exist \citep{vannoort:12,asensioramos:15a}, but for them to function properly, they require the exact knowledge of the PSF at all times during the scan, which is extremely difficult to achieve for ground-based measurements with varying seeing conditions. An extension of these methods to cope with such a time-varying PSF, to be accurately measured during the scan, is required to fully exploit the high-resolution measurements of GREGOR and future solar telescopes with larger apertures.

\begin{acknowledgements}
The 1.5-meter GREGOR solar telescope was built by a German consortium under the leadership of the Kiepenheuer-Institut f\"ur Sonnenphysik in Freiburg with the Leibniz-Institut f\"ur Astrophysik Potsdam, the Institut f\"ur Astrophysik G\"ottingen, and the Max-Planck-Institut f\"ur Sonnensystemforschung in G\"ottingen as partners, and with contributions by the Instituto de Astrof\'isica de Canarias and the Astronomical Institute of the Academy of Sciences of the Czech Republic.

Hinode is a Japanese mission developed and launched by ISAS/JAXA, collaborating with NAOJ as a domestic partner, NASA and STFC (UK) as international partners. Scientific operation of the Hinode mission is conducted by the Hinode science team organized at ISAS/JAXA. This team mainly consists of scientists from institutes in the partner countries. Support for the post-launch operation is provided by JAXA and NAOJ (Japan), STFC (U.K.), NASA, ESA, and NSC (Norway).

This work was partly supported by the BK21 plus program through the National Research Foundation (NRF) funded by the Ministry of Education of Korea.

This study is supported by the European Commissions FP7 Capacities Programme under the Grant Agreement number 312495.

The GRIS instrument was developed thanks to the support by the Spanish Ministry of Economy and Competitiveness through the project AYA2010-18029 (Solar Magnetism and Astrophysical Spectropolarimetry).
\end{acknowledgements}

\end{document}